**Title page**

# Critical Flicker Fusion Frequency As An Experience-Restricted Constraint On Visual Temporal Resolution: What Does And Does Not Change It

Natalia D. Rydzenska[1, **], Pawel J. Winklewski[2], Michał W. Błaszczyk-Niezgoda[2, 3], Anna B. Marcinkowska[1, *]

[1] Applied Cognitive Neuroscience Lab, Department of Neurophysiology, Neuropsychology, and Neuroinformatics, Medical University of Gdansk, Gdansk, Poland

[2] Department of Neurophysiology, Neuropsychology, and Neuroinformatics, Medical University of Gdansk, Gdansk, Poland

[3] Department of Neurology and Stroke, Nicolaus Copernicus Hospital, COPERNICUS PL Sp. z o. o., Gdansk, Poland

** Formerly published as Mankowska; see ORCID 0000-0001-9704-4226 for a complete publication record.

**Corresponding author**

* Anna B. Marcinkowska, Applied Cognitive Neuroscience Lab, Department of Neurophysiology, Neuropsychology, and Neuroinformatics, Medical University of Gdansk, Tuwima 15, 80-210 Gdansk, Poland. E-mail: anna.marcinkowska@gumed.edu.pl. Phone: +48 58 349 1515

**ORCID**

Natalia D. Rydzenska (Mankowska): https://orcid.org/0000-0001-9704-4226

Pawel J. Winklewski: https://orcid.org/0000-0002-1671-2163

Michał W. Błaszczyk-Niezgoda: https://orcid.org/0000-0002-6404-0568

Anna B. Marcinkowska: https://orcid.org/0000-0002-8331-208X

**Author contributions (CRediT)**

Natalia D. Rydzenska: Conceptualization, Investigation, Visualization, Writing – original draft, Writing – review & editing.

Pawel J. Winklewski: Conceptualization, Supervision, Writing – original draft, Writing – review & editing.

Michał W. Błaszczyk-Niezgoda: Investigation, Writing – review & editing.

Anna B. Marcinkowska: Conceptualization, Supervision, Project administration, Visualization, Writing – review & editing.

**Funding**

This research did not receive any specific grant from funding agencies in the public, commercial, or not-for-profit sectors.

**Competing interests**

The authors declare that they have no known competing financial interests or personal relationships that could have appeared to influence the work reported in this paper.

# Critical Flicker Fusion Frequency As An Experience-Restricted Constraint On Visual Temporal Resolution: What Does And Does Not Change It

## Abstract

Experience-dependent plasticity is fundamental to adaptive behaviour, yet the conditions under which basic sensory timing can be modified in adulthood remain poorly specified. Critical flicker fusion frequency (CFFF), the threshold at which flicker is perceived as continuous, is unusually informative here: what fails to change it is as well documented as what does. We argue that CFFF is best characterised neither as non-plastic nor as held near a physiological ceiling, but as modifiable only by a specific class of experience, not by amount. Repeated testing changes it little, and cognitive training without temporal content and incoherent-motion exposure leave it unchanged regardless of duration. By contrast, perceptual-learning paradigms that repeatedly drive early visual processing of transient signals can raise it. The evidence is graded: strongest for task-irrelevant coherent motion temporally paired with task targets, with one-year retention in a small subsample; weaker for contrast and dot-motion training; and, for flicker training, confined to amblyopic eyes starting below typical values. Three qualifications apply: samples are small; the motion and contrast paradigms were assessed with a small foveal stimulus on a macular pigment densitometer, whose equivalence with standard CFFF measurement is unestablished; and a minority of untrained controls show comparable changes. We examine whether the restriction originates at peripheral, thalamocortical, or cortical levels; available data do not adjudicate between them. Functional arguments for such a restriction are offered as rationales, not evidence. Proposed links to working-memory precision and metacognition are stated as predictions; first tests were largely negative.

## 1. Introduction

Experience-dependent neural plasticity is fundamental to adaptive behaviour, enabling organisms to optimise their interactions with dynamic environments. The discovery of critical periods originally revealed that certain neural properties can be modified by experience only during specific developmental windows (Wiesel & Hubel, 1963). However, mounting evidence demonstrates that the adult cortex retains substantial plasticity well beyond these traditional sensitive periods (Bavelier et al., 2010; Gilbert & Li, 2012). This adult plasticity manifests through multiple mechanisms, including perceptual learning-induced modifications of the primary visual cortex (Li et al., 2004, 2008), cross-modal recruitment following sensory deprivation (Lee & Whitt, 2015; Whitt et al., 2014), adaptation-induced plasticity through repeated stimulus exposure (Bharmauria et al., 2022), and metaplastic changes in excitatory–inhibitory balance (Froemke, 2015).

Importantly, neural plasticity operates within a hierarchically organised system where intrinsic neural timescales increase systematically from the sensory to the association cortex (Chaudhuri et al., 2015; Murray et al., 2014). Training modulates both intrinsic timescales (spontaneous fluctuations) and task-evoked dynamics in experience-dependent ways that contribute to improved cognitive performance (Trepka et al., 2024). This organisation is particularly evident in the prefrontal cortex, which serves as an adaptive hub that integrates past experience with present goals, bridging working memory and long-term learning.

Yet this landscape of extensive plasticity is not uniform, and its structure is better described in three tiers than in two. Some properties are set by physical and biophysical law and cannot be altered by any experiential intervention: the kinetics of phototransduction, the relation between axon geometry and conduction velocity, the energetic cost of a spike. Others are established during development and remain unchanged under ordinary conditions yet can be shifted by training that engages the specific pathway supporting them. Others again are modified readily and continuously throughout life.

We retain the familiar computing metaphor for the first and third tiers but note that its two-term version misdescribes the middle tier, which is where critical flicker fusion frequency (CFFF) lies. Hardware, in this usage, denotes properties that no intervention can alter; software denotes properties reconfigured by ordinary use. The middle tier is better captured by firmware: functionality that is fixed relative to ordinary use, that cannot be changed by running different programs on top of it, but that is nonetheless updatable through a dedicated channel. The analogy is useful here because it captures two features of the CFFF training literature simultaneously (Section 1.2): general cognitive activity and repeated testing leave the threshold essentially unchanged, whereas a restricted class of perceptual-learning paradigms can raise it, in one study durably.

We use the metaphor descriptively and state where it fails. Firmware updates are discrete, deliberate and permanent, whereas the perceptual learning at issue is graded and is not the operation of a mechanism that evolved to be updated. Nothing in the argument depends on the analogy: the anatomical claims are made directly in Section 2, where the middle tier is provisionally mapped onto thalamocortical feedforward connectivity stabilised during critical periods, and the upper tiers onto intracortical horizontal and feedback connectivity that remains modifiable in adults.

Historically, certain abilities were considered impervious to adult plasticity based on limited training studies and critical period theory (Knudsen, 2004). Recent evidence has challenged many of these assumptions: adults can acquire absolute pitch when provided with extended, gamified training (Van Hedger et al., 2019). Furthermore, pharmacological reopening of critical-period plasticity through histone deacetylase inhibition shows that the closure of learning windows reflects active molecular braking rather than irreversible constraint (Gervain et al., 2013). Finally, phoneme discrimination training produces durable perceptual gains in older adults with mild hearing loss, retained at one-month follow-up, even though benefit size varies across individuals and does not consistently generalise to everyday speech-in-noise perception (Ferguson et al., 2014). These findings suggest that many apparent "exceptions" to adult plasticity reflect insufficient training intensity or overly rigid interpretations of critical-period constraints rather than fundamental neurobiological limits (Hooks & Chen, 2007). Against this backdrop, critical flicker fusion frequency (CFFF) is an instructive case, not because it is exempt from this pattern but because its bounds are unusually sharp and unusually easy to specify. Defined as the threshold above which a flickering light is perceived as continuous (reviewed by Mankowska et al., 2021), CFFF offers a tractable probe of this middle tier: a behavioural threshold whose bounds on modification appear unusually sharp, both in the paradigms that move it and in the magnitude of change they produce. Despite decades of psychophysical research, clinical application, and implicit daily exposure to flicker, CFFF shows little or no improvement under repeated testing or cognitive training without temporal content (Bernardi et al., 2007; Görtelmeyer & Wiemann, 1982; Muth et al., 2026). This stands in sharp contrast to other capacities processed through the same early pathways: orientation discrimination, contrast detection, and spatial frequency discrimination, all processed through the primary visual cortex (V1), demonstrate robust perceptual learning (Polat, 2009; Sagi, 2011), whereas CFFF remains essentially stable within individuals across extended periods.

We return to this dissociation in Section 2.2, where we ask whether it reflects a difference between temporal and spatial properties processed in the same cortical territory. CFFF operates at timescales (typically 30–60 Hz) that span peripheral retinal processing and early cortical integration (Mankowska et al., 2022), which raises the question of whether the restriction on its modification reflects peripheral limitations, thalamocortical transmission dynamics, or actively maintained cortical temporal filtering.

We do not claim that CFFF marks a boundary between plastic and non-plastic systems. Our claim is narrower: CFFF is a property whose modification is gated by the class of experience rather than by its amount. One possible functional reason for such gating is that CFFF may index a "perceptual clock" providing a stable temporal reference frame for higher-order processing (Section 3.1). If temporal resolution shifted readily with ordinary experience, learned temporal associations would need continual recalibration, and the integration of discrete sensory events within brief windows (~50–300 ms) into unified percepts such as audio-visual speech could become less reliable (Powers et al., 2009; Wallace & Stevenson, 2014). We offer this as a rationale, not as evidence.

### 1.1 Defining an experience-restricted constraint

Because experience-restricted constraint is the central construct of this review, we define it operationally. We treat a perceptual property as experience-restricted when it satisfies three conditions: (1) high within-individual stability under standardised measurement, with no systematic change under repeated measurement; (2) insensitivity to experience falling outside a delimited class, such that increasing the duration, difficulty or engagement of training outside that class does not produce change;

and (3) consolidation of whatever change the admitted class does produce, so that gains persist after training ends rather than decaying as state effects do.

These three conditions do the work that a claim of non-plasticity cannot. Conditions (1) and (2) are what a stability claim amounts to once stated carefully: the property neither drifts nor responds to the amount of experience. Condition (3) separates genuine modification from arousal-driven fluctuation. A property meeting all three is not fixed but gated, and the theoretically interesting question becomes what the gate admits. We show in Section 1.2 that CFFF satisfies conditions (1) and (2), and that condition (3) is supported by the only relevant retention data available, which come from three participants in a single study.

We do not build a mechanistic condition into the construct. Attribution of the restriction to critical-period-stabilised thalamocortical feedforward circuitry is an interpretive hypothesis for which no study has yet linked human CFFF to thalamocortical signatures. It is developed in Section 2 and returned to as a neuroimaging prediction in Section 6, but it is not used to define the construct or to support the central claim.

We also clarify terminology. We use critical flicker fusion frequency (CFFF) throughout as the construct of interest, representing the maximum temporal frequency resolvable as flicker, and treat the operational variants (ascending and descending methods of limits, method of constant stimuli, staircase) as alternative estimators of that construct, following Muth et al. (2023) and Eisen-Enosh et al. (2017). One divergence, however, cannot be absorbed under that assumption and is material to the argument. The motion- and contrast-training studies reporting change measured the threshold on a macular pigment densitometer in Maxwellian view with a small (1°) foveal spot: a blue–green stimulus that was not equiluminant, so that the percept was mainly luminance flicker (Seitz et al., 2006), or a single green (543 nm) light (Zhou et al., 2016). Thresholds obtained in this way were about 18–21 Hz (Zhou et al., 2016). The studies establishing stability and reliability, and the flicker-training study in amblyopia (Eisen-Enosh et al., 2023), used LED-based stimuli with thresholds of about 23–50 Hz. Whether the two procedures estimate the same underlying limit is unestablished, and we flag the divergence wherever the two literatures are compared.

### 1.2 The empirical basis: what changes CFFF and what does not

The claim that CFFF is experience-restricted rests on an asymmetry in the literature: what fails to alter the threshold is documented about as well as what alters it. We summarise both sides here; the full evidence base is tabulated in Supplementary Table S1 (training and intervention studies) and Supplementary Table S2 (stability and state modulation).

Stability. Under standardised conditions, baseline thresholds are highly stable within individuals. A single-participant series of daily measurements across 38 days under controlled lighting yielded 44.0 +/- 1.7 Hz, a coefficient of variation of approximately 4% (Muth et al., 2023), and an independent single-participant series across 85 consecutive days showed approximately 5% variation (Muth et al., 2026). In a cohort of 49 participants measured three times at intervals of at least 24 hours, approximately 80% of total variance was between participants; about 10% was associated with session, reflecting a systematic rise of 0.9–1.3 Hz per session, and 4–9% was residual within-participant variance (Haarlem et al., 2024). Within-method test-retest correlations are high for the staircase ($r = 0.95$) and constant-stimuli ($r = 0.946$) procedures and moderate for the method of limits ($r = 0.72$), with mean relative test-retest differences of approximately 3%, 3% and 6% respectively (Eisen-Enosh et al., 2017); the frequently cited values of $r = 0.92$ and +/- 3.6 Hz limits of agreement from the same study describe agreement between methods rather than test-retest reliability. Two qualifications apply: the longitudinal evidence rests substantially on two single-participant series, and stability observed without intervention establishes the absence of drift, not the absence of trainability.

What does not change CFFF. Repeated measurement produces a familiarisation effect of approximately 3% between the first and second administration, which saturates thereafter (Bernardi et al., 2007), and untrained control participants measured daily across nine days rose by approximately 5% (range 3-9%; Seitz et al., 2006). Saturation is not universal: across three sessions, thresholds rose by about 1.6 Hz per session in women and about 0.4 Hz in men, and the change was significant only in women (Haarlem et al., 2024). An active Sudoku control produced no change over eight sessions (Zhou et al., 2016). Exposure to the flickering stimulus without coherent motion likewise produced no reliable change: participants trained with dot displays at 0% coherence improved by approximately 4% on average (range -5 to 13%), not significantly different from a flicker-only control (Seitz et al., 2006). Five sessions of flicker-based temporal training left normally sighted observers unchanged (30.34 +/- 0.5 to 30.7 +/- 0.7 Hz, p = 0.92; Eisen-Enosh et al., 2023). A report of a 15% increase after 30 days of yoga (Vani et al., 1997) is sometimes read as a counterexample, but the allocation was not randomised and the group by assessment interaction was not significant (F = 0.1), so the analysis does not support a training-specific effect.

What does change CFFF. Gains have been reported for a restricted class of paradigms, with evidence of differing strength. Participants performed a central rapid serial visual presentation task requiring identification of two target letters while coherent directional motion, presented peripherally at a contrast below their detection threshold and therefore irrelevant to the task, was temporally paired with the targets. Across nine daily sessions the threshold rose by approximately 30% in five participants, each showing a significant association between threshold and training day (r-squared 0.77-0.94), and a separate group of three tested only before and after training rose by 22% (range 18-27%) (Seitz et al., 2005, 2006). Three of the five trained participants retested one year later had changed by less than 1% from their post-training level. In a later study, eight sessions of contrast-sensitivity training raised the threshold by 1.36 Hz and directional dot-motion training by 0.74 Hz, although the group by session interaction was only a trend (p = 0.08) and only the contrast-training group differed significantly from the Sudoku control (Zhou et al., 2016). In amblyopia, where the amblyopic eye begins well below the fellow eye, five sessions of flicker-based training raised the threshold from 22.8 +/- 3.4 to 26.7 +/- 3.15 Hz, an increase of 17% (Eisen-Enosh et al., 2023).

Taken together, these studies define the admitted class only provisionally. What they share is perceptual learning that repeatedly drives early visual processing of transient signals (motion, contrast or flicker). Within that class the evidence is graded: it is strongest for task-irrelevant coherent motion temporally paired with task targets, weaker and partly non-significant for contrast and dot-motion training, and, for flicker training, confined to amblyopic eyes starting below typical values. Class membership is therefore not sufficient for change: flicker training did not alter the threshold of normally sighted observers. Exposure to the same dot displays without coherent motion was also ineffective, which suggests that transient stimulation without a coherent, task-paired signal is not enough. Whether the operative factor is the transient signal, its temporal pairing with a task-relevant event, or the headroom available in the trained system cannot be decided from studies of this size. Outside the domain of experience, four months of macular carotenoid supplementation produced a small increase relative to placebo in a randomised, double-blind trial (26.6 to 27.3 Hz in the active groups; Bovier et al., 2014), which shows that the threshold of normally sighted adults is not closed to physiological intervention.

Three qualifications. First, every training study cited above used between three and ten participants per group, and several analyses were one-tailed; the effect sizes should be treated as provisional. Second, the motion and contrast paradigms that raise the threshold in normally sighted adults were assessed with a small foveal stimulus on a macular pigment densitometer, and construct equivalence with standard LED-based CFFF is unestablished (Section 1.1). Third, and least easily accommodated, the distribution of change among untrained control participants is bimodal: of 19 controls pooled across conditions, 14 changed by 5% or less but five increased by 10-17%, and only those five also improved in motion-direction sensitivity (Seitz et al., 2006). If that covariation reflects incidental exposure to the admitted class of experience during testing, the account holds and the gate is easier to pass than the design intended; if it does not, part of the effect is unexplained.

This last observation is also why we do not adopt a fixed percentage threshold for meaningful change. A 10% criterion would classify five untrained controls as having been trained while excluding the gain from contrast training that Zhou et al. (2016) found to differ significantly from an active control. We therefore state the predictions below as control-referenced comparisons with pre-specified retention intervals rather than as absolute cut-offs, and treat changes below approximately 6% as within the range of measurement and familiarisation effects.

Five predictions follow. All are stated prospectively; the evidence above motivates them rather than confirming them.

- Class restriction. Training that does not drive early visual processing of transient signals will produce no change relative to an active control, regardless of dose. Falsified by a replicated, control-referenced gain from such a paradigm.

- Pairing dependence. Within the admitted class, disrupting the temporal pairing between the motion signal and the task target will abolish the gain. Partially supported by the 0%-coherence and unpaired control conditions of Seitz et al. (2006); requires replication with adequate samples.

- Headroom dependence. The magnitude of gain will scale inversely with baseline threshold. Consistent with the contrast between amblyopic and normally sighted observers (Eisen-Enosh et al., 2023) but confounded with population, and untested within samples.

- Consolidation. Gains obtained through the admitted class will persist for months after training ends. Supported at one year in three participants (Seitz et al., 2006); requires a pre-specified retention design.

- Construct generality. Gains obtained with the small-field densitometer stimulus will replicate when the threshold is measured with a standard LED-based procedure. Untested. A failure would localise the effect to the measurement procedure rather than to visual temporal resolution.

We regard the fifth prediction as the most consequential, because the evidence for change in normally sighted adults currently rests on a single measurement procedure.

The framework also bears on individual differences in cognition. Visual working memory and temporal resolution interact bidirectionally (Pan, 2020), and working-memory uncertainty is jointly encoded with content in probabilistic neural codes (Li et al., 2021; van den Berg et al., 2017). If CFFF reflects a stable individual constraint on temporal processing speed, it may serve as a trait marker predicting the precision of higher cognitive functions – a prediction we frame as testable rather than established. In the remainder of this review we develop the hierarchical architecture of plasticity (Section 2), functional and computational rationales for restricting its modification (Section 3), CFFF as an individual-difference variable (Section 4), implications for theoretical frameworks (Section 5), and future directions (Section 6).

# 2. The Hierarchical Architecture of Neural Plasticity

### 2.1 Levels of Plasticity in Visual Processing

Neural plasticity in the visual system can be conceptualised as operating at multiple hierarchical levels, each defined by two criteria: the degree of experience-dependent modifiability in adulthood, and the developmental timing at which plasticity becomes constrained (Figure 1). Levels are distinguished by their primary anatomical substrate, the molecular mechanisms governing their stability, and the timescale over which modifications, where possible, occur. Within this framework CFFF indexes Level 1, the gated tier: a property whose modification is restricted to a narrow class of experience and whose expression is modulated by physiological state.

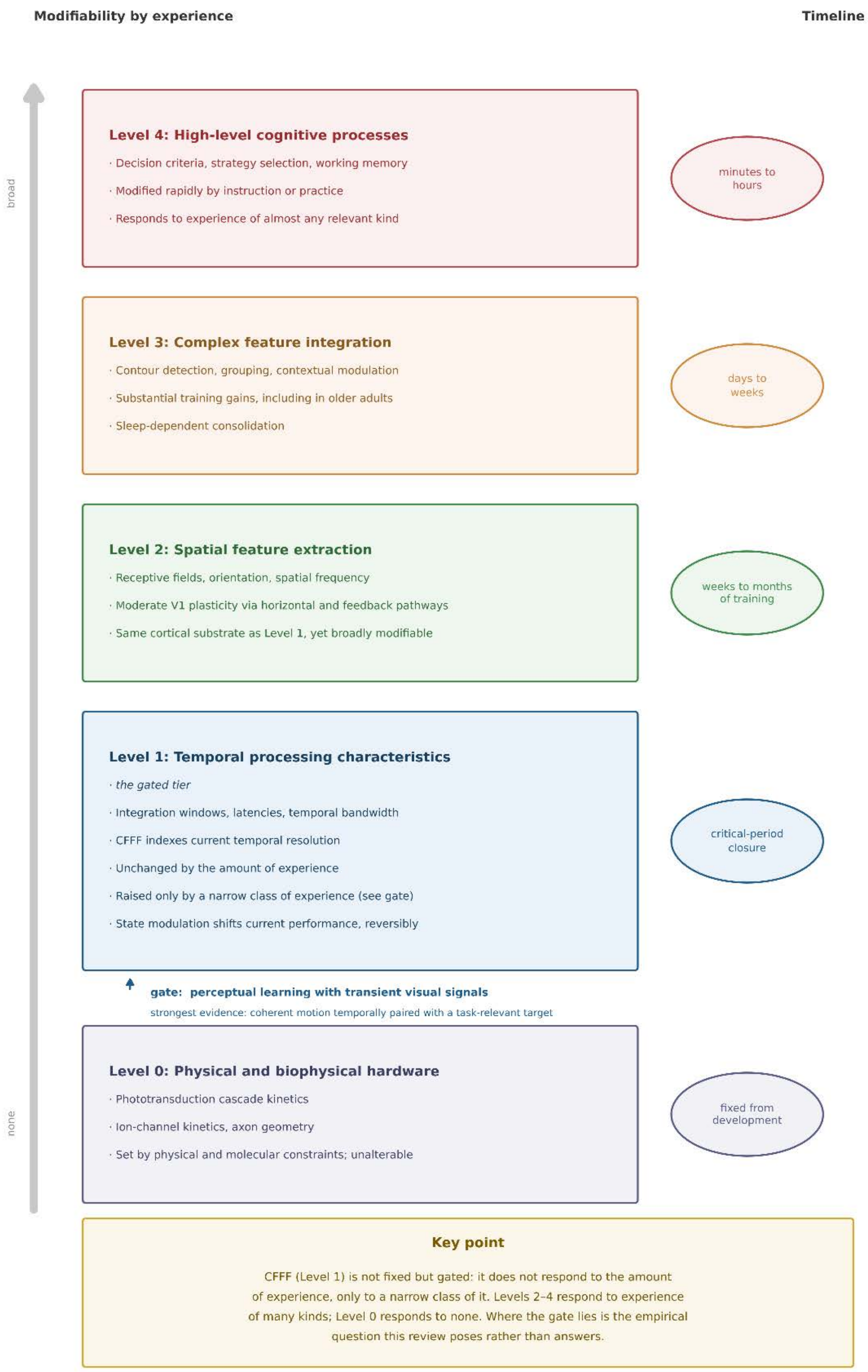

Figure 1. Hierarchical architecture of experience-dependent modifiability in visual processing. Five processing levels ordered by how broadly each responds to experience in adulthood (left axis) and by the developmental timing at which that responsiveness becomes constrained (right axis). Level 0 is unalterable by any intervention; Levels 2-4 respond to experience of many kinds; Level 1, which CFFF indexes, is gated - it does not respond to the amount of experience but can be raised by a narrow class of it, provisionally identified as perceptual learning that drives early visual processing of transient signals, with the strongest evidence for coherent motion temporally paired with a task-relevant target (Section 1.2). State-dependent modulation at Level 1 shifts current performance reversibly and is distinguished from gated modification. Where the gate lies is the empirical question this review poses rather than answers.

### 2.1.1 Level 0 – Biophysical hardware

At the most fundamental level, processing is governed by physical and biophysical constraints determined by molecular structure and thermodynamic principles. These properties are established during development and remain immutable in adulthood – they cannot be modified by experience, training, or pharmacological intervention. The phototransduction cascade is canonical: photon absorption by rhodopsin initiates an amplification cascade whose kinetics are set by the rate constants of G-protein coupling, phosphodiesterase activation, and calcium feedback (Burns & Baylor, 2001; Saari, 2000). Axonal conduction velocity is constrained by membrane capacitance and axial resistance (Debanne et al., 2011; Rushton, 1951), although myelination itself remains partly activity-dependent in adulthood (Pajevic et al., 2014) and is therefore not wholly a Level 0 property.

### 2.1.2 Level 1 – Temporal processing characteristics: the firmware tier

In the terms introduced in Section 1, Level 0 corresponds to hardware and Level 1 to firmware. The distinction between them is that Level 0 properties cannot be modified by any experience, whereas Level 1 properties can be modified, but only through a restricted channel. This difference is the substantive content of the present account and is developed empirically in Sections 1.2 and 2.2. Level 1 encompasses fundamental temporal processing characteristics determined by neural circuit architecture established during developmentally regulated critical periods. These properties set an individual's current limit on temporal resolution, a value that ordinary experience does not shift. Unlike Level 0 constraints, they are plausibly set by experience-dependent circuit maturation, in which excitatory-inhibitory balance, glial programs and myelination have all been implicated (Hensch, 2005; Xin et al., 2024).

We take CFFF as the most accessible behavioural index of Level 1. Peak CFFF thresholds show substantial between-individual variation spanning a 30 Hz range, while remaining stable within individuals across extended periods (Haarlem et al., 2024). Temporal integration windows average approximately 65 ms in adults (Wutz et al., 2016) and reach adult-like values by age 5–7 (Freschl et al., 2019), with peak alpha frequency tracking the maturation of visual temporal processing across development (Freschl et al., 2022). Meanwhile, temporal frequency tuning peaks at 15–20 Hz under photopic conditions with a cut-off near 60 Hz (Kelly, 1962; Tyler, 1985). The within-individual stability and cross-method agreement of CFFF are detailed in Section 1.2. Animal work shows that thalamocortical feedforward connections are stabilised during critical periods through parvalbumin-interneuron maturation and perineuronal-net deposition (Hensch, 2005; Ribic et al., 2019). Whether the critical period for visual temporal processing in humans closes earlier than that for spatial properties has not, to our knowledge, been established, and we treat the link to CFFF as a hypothesis (Section 1.1).

**State-dependent modulation of Level 1.** Level 1 properties must be distinguished from their state-dependent expression. The limit set during development is modulated by physiological state without itself being shifted (Figure 1). Arousal enhances temporal resolution through increased thalamocortical firing and neuromodulatory mechanisms (Chen et al., 2015; Mather & Sutherland, 2011). Conversely, fatigue and sleep deprivation reduce it through time-on-task effects (Muth et al., 2023).

Acute physiological stressors illustrate this principle especially clearly: hyperbaric nitrogen narcosis transiently reduces CFFF, whereas hyperoxia and altered gravitational loading shift it in dose-dependent and direction-specific ways. These effects largely reverse when normal conditions are restored, although not always within the measurement window: after a 33 m fresh-water dive, CFFF remained below baseline 30 min after surfacing (Balestra et al., 2012, 2018). Reversal over hours distinguishes such state effects from the consolidated change required by condition (3) in Section 1.1. State-dependent CFFF changes should therefore not be read as evidence of modification of the limit itself; equally, their reversibility is consistent with, but does not establish, a stable underlying limit.

### 2.1.3 Level 2 – Spatial feature extraction: the onset of software-like plasticity

At Level 2, spatial feature extraction shows moderate adult plasticity. Receptive-field properties, orientation tuning, and spatial-frequency selectivity, which are all instantiated within V1, can be modified through perceptual learning (Li et al., 2004, 2008; Schoups et al., 2001). This process involves morphological changes in horizontal connections (axon-collateral sprouting and pruning; Chen et al., 2011; Yamahachi et al., 2009). It also includes tuning-curve sharpening through feedforward and intracortical refinement (Teich & Qian, 2010), operating as a two-phase process of V1 activation change followed by white-matter connectivity modification (Astorga et al., 2022; Lu & Dosher, 2022).

Levels 1 and 2 are both expressed in V1 yet differ in how they respond to experience. One candidate explanation, examined in Section 2.2, is anatomical: Level 2 plasticity is mediated by modifiable horizontal and feedback pathways, whereas Level 1 temporal properties may depend more on critical-period-stabilised thalamocortical feedforward connections (Espinosa & Stryker, 2012; Hooks & Chen, 2006).

2.1.4 Level 3 – Complex feature integration

Level 3 encompasses complex feature integration, including contour detection, perceptual grouping, and contextual modulation. This level shows high plasticity, with training producing substantial improvement over days to weeks even in older adults (Gilbert et al., 2009; McKendrick & Battista, 2013; Sagi & Tanne, 1994). These improvements are supported by sleep-dependent consolidation and reactivation-based learning (Kondat et al., 2024; Stickgold & Walker, 2007). Perceptual grouping transfers learning to grouped stimuli without voluntary attention (Mastropasqua & Turatto, 2013), utilising mechanisms that span the occipital, fusiform, and parietal cortex (Hua et al., 2023). Furthermore, texture-discrimination learning remains feature-specific while resisting local adaptation (Censor & Sagi, 2009; Kahalani-Hodedany et al., 2024). Contextual modulation depends on VIP-mediated disinhibition of SOM interneurons (Keller et al., 2020) and on interlaminar signal flow from feedback/horizontal layers (S. Zhu et al., 2024).

### 2.1.5 Level 4 – High-level cognitive processes

Level 4 comprises decision criteria, strategy selection, and working-memory operations. These processes show extreme plasticity and can be modified rapidly through instruction or practice. Decision criteria are flexibly adjusted with attentional state and task demands (Lee et al., 2023), perceptual decisions combine reinforcement learning and evidence accumulation (Urai et al., 2019). Similarly, working-memory capacity shows learning-induced gains (Conci et al., 2023), and cognitive flexibility is supported by complementary dopaminergic habit- and goal-based policies (Miederer et al., 2025; Sandbrink & Summerfield, 2024). These processes are distinguished from lower-level plasticity by their broad generalisability and rapid modifiability (Lu & Dosher, 2022).

## 2.2 Dissociation Within Primary Visual Cortex

The contrast between temporal and spatial properties within V1 motivates the anatomical hypothesis of the present account (the behavioural evidence is summarised in Figure 2). V1 exhibits robust plasticity for spatial properties, including the sharpening of orientation tuning (Schoups et al., 2001), receptive-field modification (Fu et al., 2002; Yao & Dan, 2001), and enhanced contextual integration (Li et al., 2004, 2008). Comparable plasticity of temporal integration windows in V1 has not, to our knowledge, been reported, although it has rarely been sought. This cannot be explained by a simple "plastic vs. non-plastic cortex" dichotomy; rather, different computational properties within the same area are subject to different constraints. Adult V1 plasticity primarily involves horizontal and feedback connections, while feedforward thalamocortical connections that underwent critical-period plasticity during development remain relatively fixed (Espinosa & Stryker, 2012; Hooks & Chen, 2006).

The fact that peripheral sensory deprivation can transiently reactivate thalamocortical plasticity through GluN2B re-expression and silent-synapse mechanisms (Chung et al., 2017) suggests, in rodent somatosensory cortex, that this stability is actively maintained rather than passively lost. Critical-period plasticity specifically targets thalamocortical inputs onto parvalbumin interneurons via synaptic adhesion molecules such as SynCAM 1 (Ribic et al., 2019).

Temporal integration properties may thus be tied to feedforward streams and fundamental cellular properties (membrane time constants, synaptic kinetics) set during development. Short-term synaptic depression at feedforward L4 to L2/3 synapses generates rapid adaptation with excitatory transmission recovering within a few seconds (Lombardi et al., 2023). Because the feedforward laminar circuit is broadly conserved across sensory cortices, this finding from barrel cortex may generalise to visual cortex, but this has not been tested. Spatial properties, by contrast, depend on lateral and feedback connectivity that retains lifelong plasticity, and feedforward and feedback interactions use distinct population activity patterns (Semedo et al., 2022). If CFFF depends mainly on early feedforward temporal processing, this architecture would account for its restricted modifiability, whereas tasks dependent on lateral integration (e.g., contour detection) improve readily. The hypothesis is testable (Section 6.4) but untested.

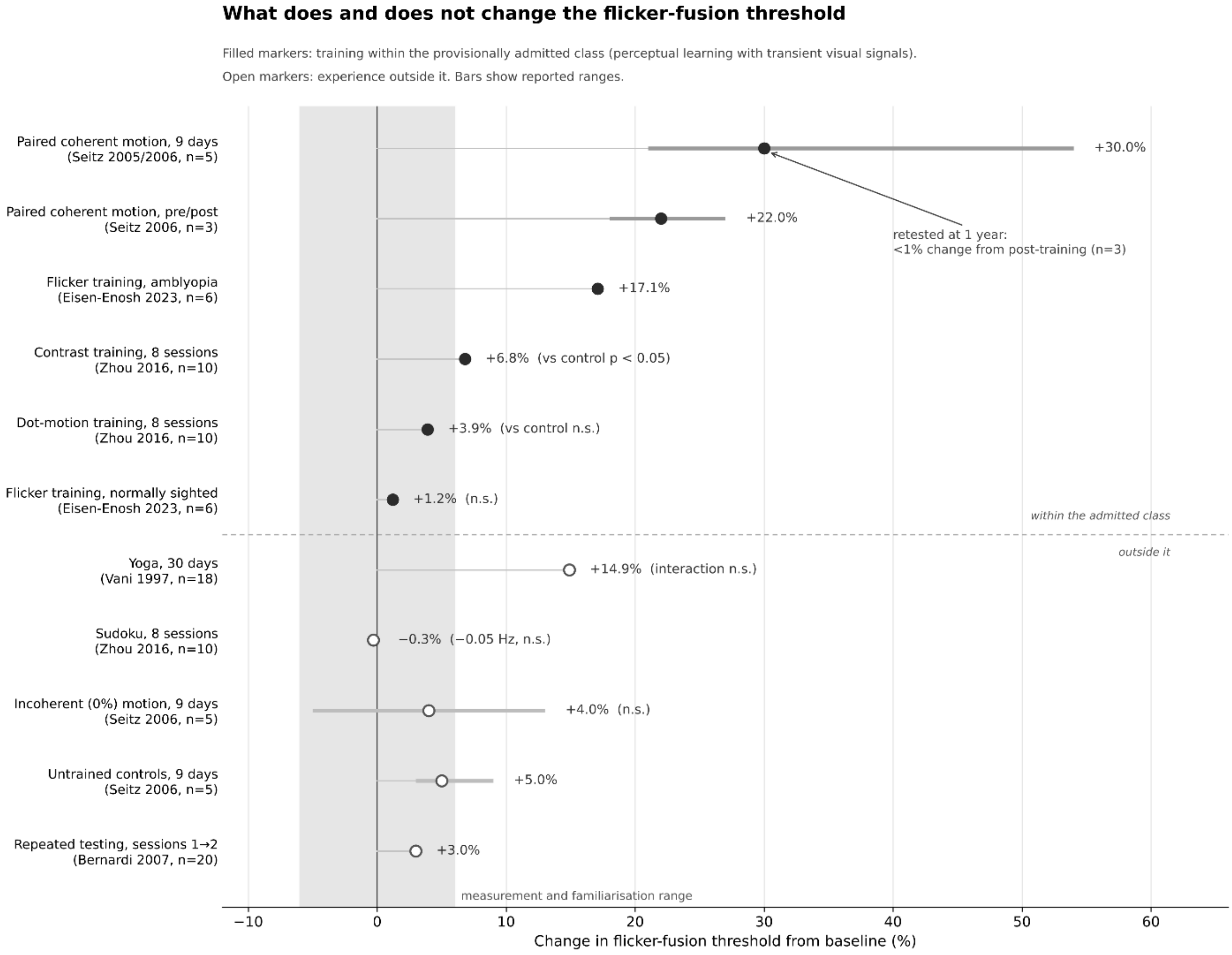

Figure 2. What does and does not change the flicker-fusion threshold. Reported percentage change from baseline for every training, control and intervention condition reviewed in Section 1.2 and tabulated in Supplementary Table S1. Points are group means and horizontal bars show reported ranges where available; the shaded band marks the range attributable to measurement and familiarisation effects (Bernardi et al., 2007; Eisen-Enosh et al., 2017). Filled markers denote training within the provisionally admitted class (Section 1.2), including flicker training in normally sighted observers, which produced no change; open markers denote experience outside it. Sample sizes are given with each condition and are small throughout; the contrast-training condition of Zhou et al. (2016) lies just outside the shaded band and the dot-motion condition within it, and the yoga result of Vani et al. (1997) is plotted with its non-significant group-by-assessment interaction noted. Values obtained with a small-field densitometer stimulus (Seitz et al., 2006; Zhou et al., 2016) and with LED-based stimuli (Eisen-Enosh et al., 2023) are shown together as percentage change, but their construct equivalence is unestablished (Section 1.1).

### 2.3 Evidence from Critical Period Research

Critical-period closure for different visual properties is not uniform but follows a hierarchical pattern. Ocular-dominance plasticity in V1 peaks around postnatal day 28 and ends by P32–36 in mice (Gordon & Stryker, 1996; Levelt & Hübener, 2012), though some plasticity persists in layers 2/3 and 5 (Daw et al., 1992). Closure is actively regulated by the maturation of inhibitory circuits, particularly parvalbumin-positive (PV+) interneurons (Hensch, 2005; Hensch et al., 1998), which regulate plasticity through perineuronal-net interactions, gain control, and network synchronisation (Rupert & Shea, 2022). Chemogenetic silencing of PV+ interneurons reinstates critical-period plasticity in the adult cortex (Cisneros-Franco & de Villers-Sidani, 2019; Lensjø et al., 2017). Furthermore, PV+ maturation stabilises thalamocortical connections via SynCAM 1 (Ribic et al., 2019) while horizontal connections retain adult plasticity (Kuhlman et al., 2013). Adult ocular-dominance plasticity in both V1 and dLGN depends on thalamic GABAergic inhibition (Qin et al., 2023), indicating that different forms of plasticity are independently regulated even within one circuit. If temporal integration depends on thalamocortical dynamics stabilised by PV+ maturation and thalamic inhibition, while the intracortical horizontal connections that mediate spatial integration remain plastic, this architecture could underlie the restricted modifiability of CFFF. The evidence for this link is indirect and comes almost entirely from rodent ocular-dominance paradigms.

### 2.4 Measurement Considerations and Stimulus Parameters

CFFF is not a single fixed number but a threshold that depends systematically on stimulus parameters – luminance, field size, retinal eccentricity, wavelength, and contrast all shift measured values (Kelly, 1962; Mankowska et al., 2021; Muth et al., 2023; Tyler, 1985). Treating CFFF as a unitary construct is therefore justified only when measurement conditions are specified and held constant. Two points matter for the present argument. First, the high test–retest reliability that grounds the stability claim (Section 1.2) is itself conditional on standardised conditions. Specifically, reliability is highest for staircase and constant-stimuli paradigms under controlled luminance and viewing geometry (Eisen-Enosh et al., 2017), and degrades when conditions vary. Second, several confounders of CFFF (e.g. diurnal phase, fatigue, arousal, stimulant exposure) are precisely the state-dependent modulators discussed in Section 2.1.2. Their predictable, reversible influence (Muth et al., 2023, 2026) is consistent with, rather than contrary to, a stable underlying limit. The stability we attribute to CFFF is thus a property of the construct measured under controlled conditions, not a claim that raw values are invariant across measurement contexts.

### 2.5 Neural correlates of CFFF

Although CFFF is a behavioural threshold, converging neurophysiological evidence links it to magnocellular and early cortical processing efficiency. Flicker sensitivity correlates with physiological indices of magnocellular pathway function (Brown et al., 2018). Moreover, inhibitory continuous theta-burst stimulation over the left inferior parietal lobule, but not V1 or right IPL, alters the flicker-fusion threshold. This interaction identifies a distributed cortical network in which higher-order associative cortex contributes to the conscious readout of temporal information (Nardella et al., 2014). In clinical populations characterised by oscillatory slowing, such as hepatic encephalopathy, CFFF and individual alpha frequency decline in parallel (Butz et al., 2013). These correlates are consistent with CFFF indexing early temporal processing whose conscious readout engages higher cortical areas, but they are correlational, and the parietal result rests on a single stimulation study.

## 3. Why Might Modification of Temporal Resolution Be Restricted? Functional and Computational Rationales

### 3.1 The Perceptual Clock Hypothesis

The arguments in this section are rationales for why a restriction of the kind described in Section 1.2 might be adaptive; none of them is evidence that the restriction exists. The section follows a deliberate two-step inferential structure. In Step 1 (Section 3.1.1) we establish, on grounds independent of CFFF, why a stable temporal reference is functionally necessary. Only in Step 2 (Section 3.1.2) do we introduce CFFF as a candidate marker and evaluate its empirical stability against the prediction generated in Step 1.

### 3.1.1 Theoretical foundations: why temporal resolution requires a stable oscillatory reference

Accurate perception of the temporal structure of events – motion tracking, audiovisual synchronisation, speech comprehension – demands that the nervous system maintain a reliable internal reference for encoding intervals. Without such a reference, cross-modal binding would become indeterminate and perceptual coherence would degrade. Treisman (1963) first proposed that an endogenous pacemaker generates rhythmic pulses serving as the units of subjective time: an accumulator counts pulses, and the tally constitutes perceived duration. The model makes an architectural prediction: temporal resolution is constrained by the pacemaker's period, with a faster pacemaker yielding finer sampling. Crucially, the pacemaker frequency must be stable within an individual, since otherwise the same physical interval would yield inconsistent estimates, undermining the clock's utility. Treisman et al. (1992) formalised this with temporal-oscillator models requiring a stable reference frequency.

Treisman et al. (1994) identified cortical alpha-band oscillations (8–13 Hz) as the most plausible pacemaker substrate, linking the abstract clock to a measurable signal, namely the individual alpha frequency (IAF). Two decades of work have substantiated this link. Samaha and Postle (2015) showed, using a two-flash fusion paradigm, that individuals with higher IAF discriminate two flashes at shorter interstimulus intervals. This indicates a relationship between an endogenous oscillatory rhythm and the grain of visual temporal perception – the pattern predicted if IAF represents the clock rate. Alpha oscillations fragment continuous input into discrete epochs, with the alpha phase determining whether two proximate stimuli are perceived as simultaneous or distinct (Samaha & Romei, 2024; VanRullen, 2016).

These windows have direct consequences for multisensory integration. The temporal binding window (TBW), defined as the epoch within which cross-modal stimuli are integrated (Powers et al., 2009; Wallace & Stevenson, 2014), depends on IAF. Cecere et al. (2015) found that faster alpha rhythms predicted narrower binding windows. When the alpha frequency was experimentally driven above or below each participant's IAF using transcranial alternating current stimulation, the TBW shifted predictably, where a frequency faster-than-IAF narrowed it and slower widened it. This manipulation suggests that alpha oscillations constrain, and do not merely correlate with, temporal binding precision, although the association between alpha frequency and temporal resolution has not been consistent across studies (Samaha & Romei, 2024).

Clinical evidence provides independent support. In autism spectrum disorder the TBW is characteristically extended, meaning that paired audiovisual stimuli are bound over longer asynchronies than in neurotypical controls (Foss-Feig et al., 2010; Stevenson et al., 2014). The temporal binding deficit hypothesis posits that timing disruptions are a core feature of autism (Brock et al., 2002), with degraded temporal precision cascading upward to affect multisensory integration, speech, and social communication; the binding mechanism remains operative but its precision is degraded, requiring compensatory window enlargement (Wallace & Stevenson, 2014).

The clinical phenotype of unstable temporal binding thus illustrates, by contrast, the functional consequences of a well-calibrated clock, and disruptions to temporal structure compromise sequential perceptual stability (Fornaciai & Park, 2018; Pascucci et al., 2023). The functional stakes of clock stability are illustrated in the lower band of Figure 3, where a stable clock rate supports reliable binding, whereas a variable rate would cascade into binding and coordination failures.

Together, these converging lines – internal-clock theory, alpha-oscillation research, multisensory binding with causal neurostimulation, and clinical populations – provide an independently motivated case for the necessity of a stable temporal reference. This case does not depend on any specific psychophysical marker but instead derives from the functional architecture of temporal processing itself. Ultimately, this architecture generates a testable prediction: if visual temporal resolution is governed by a stable oscillatory pacemaker, psychophysical markers of that resolution should show high intra-individual stability combined with substantial inter-individual variability.

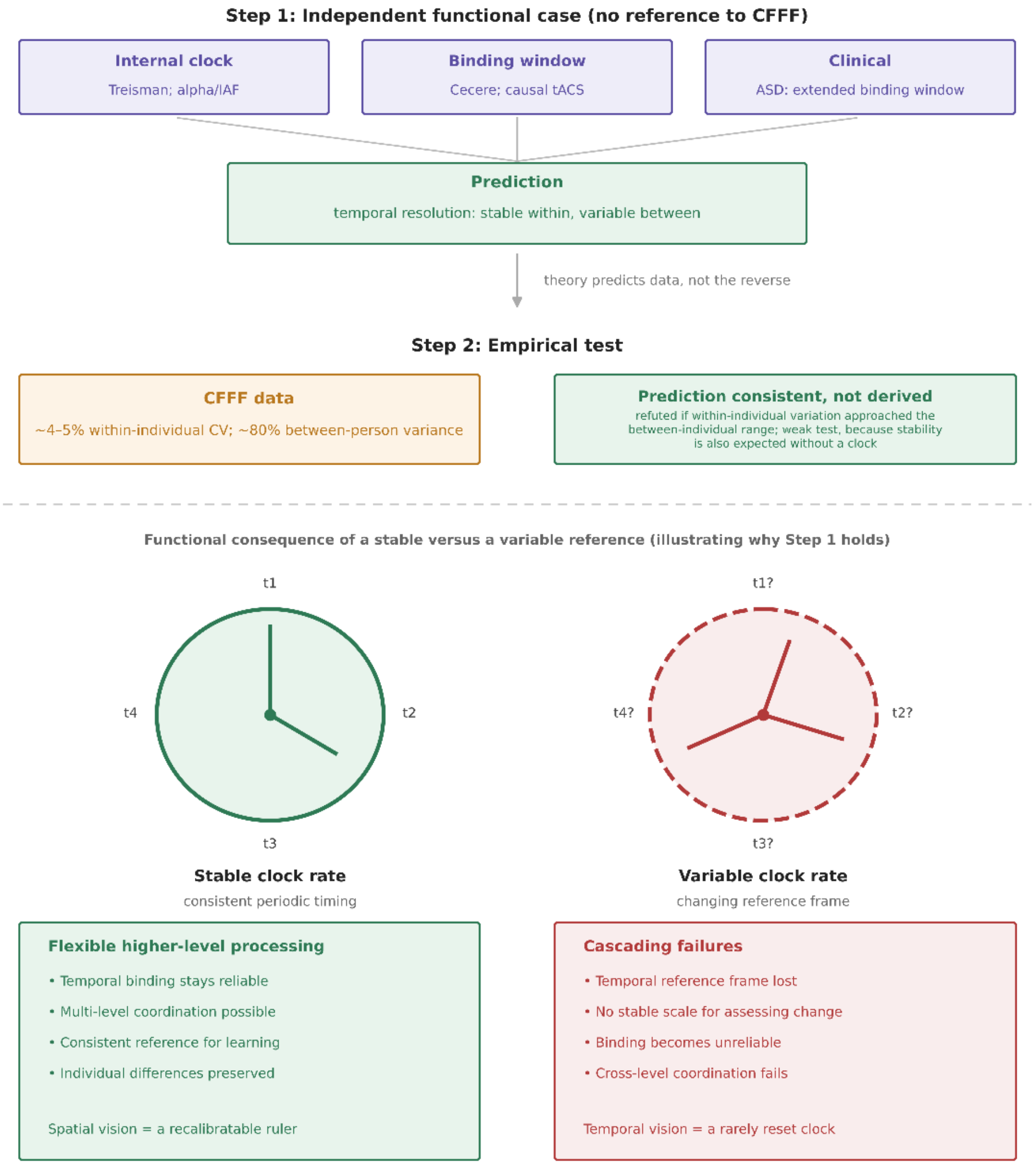

Figure 3. From functional necessity to empirical test: the logical structure of the perceptual clock hypothesis. Upper band: the two-step inferential structure. Independent functional evidence (internal clock theory, alpha and IAF, multisensory binding with causal tACS, and clinical populations) generates the prediction that temporal resolution is stable within and variable between individuals (Step 1); CFFF data are then evaluated against that prediction (Step 2). Inference runs one way - the theory predicts the data rather than being derived from them - so the argument is non-circular and falsifiable: the prediction would be refuted if within-individual variation approached the between-individual range. Because stability is also expected without a clock, the test is weak. Lower band: the functional consequence illustrated by the clock metaphor. A stable reference rate (left) supports reliable temporal binding, whereas a variable rate (right) would produce cascading binding and coordination failures. The clock illustrates why a stable reference is functionally useful (Step 1); it is not itself evidence for the hypothesis, and it does not imply that the rate is unalterable.

### 3.1.2 CFFF as a candidate marker: testing the stability prediction

CFFF provides a well-established, operationally straightforward measure against which the foregoing prediction can be evaluated. Importantly, CFFF was developed and validated independently of the internal-clock hypothesis. Its original applications were in ophthalmology, psychopharmacology, and occupational health as a marker of CNS arousal and cortical efficiency (Hindmarch, 1982; Mankowska et al., 2021). Consequently, the measure is not theoretically committed to the clock framework, and its properties can be evaluated against the clock model's predictions rather than assumed to support them.

The empirical data are consistent with this prediction. As detailed in Section 1.2, CFFF shows high inter-individual variability spanning a 30 Hz range with a 95% prediction interval of about 21 Hz. This variability is accompanied by low intra-individual fluctuation (~4–5% CV; about 80% of variance between individuals and 4–9% residual within-individual variance), with test–retest correlations of r = 0.67–0.95 over short intervals (Eisen-Enosh et al., 2017; Görtelmeyer & Wiemann, 1982) and stable single-participant series over one to three months (Muth et al., 2023, 2026). This is the signature expected if CFFF reflects a trait-like property constrained by an individually characteristic temporal architecture.

The stability of CFFF contrasts with the plasticity of other temporal measures. The multisensory TBW can be narrowed by approximately 40% through training (McGovern et al., 2022; Powers et al., 2009), and temporal recalibration occurs rapidly through brief audiovisual asynchrony exposure (Fujisaki et al., 2004; Vroomen et al., 2004). However, these effects are usually interpreted as adjustments in decision criteria and binding thresholds rather than changes in the underlying resolution limit; the studies did not measure CFFF, so this interpretation cannot be checked directly. The modest ~3% learning effect on repeated CFFF testing (Bernardi et al., 2007; Muth et al., 2023) is usually attributed to familiarisation rather than improved resolution, and values plateau thereafter.

A physiological observation is consistent with this interpretation: in hepatic encephalopathy, a condition of progressive cortical oscillatory slowing, both IAF and CFFF decline in parallel (Butz et al., 2013). Chronic-recording work further shows that temporal codes carry more stable visual representations than firing rates across 15-day periods (H. Zhu et al., 2025), and that neural stability at early latencies predicts perceptual behaviour (Baror et al., 2024). These findings suggest temporal dynamics are constrained by intrinsic circuit properties rather than flexible strategy. Thus, CFFF does not constitute the perceptual clock but instead indexes the output of a temporally stable architecture whose necessity is independently motivated above.

Two gaps should be stated. The clock models concern alpha-band rhythms (~8–13 Hz), whereas CFFF lies at roughly 30–45 Hz, so CFFF cannot be the clock rate itself and any link between them must be indirect. Moreover, the evidence relating alpha frequency to temporal resolution comes from two-flash and audiovisual paradigms, not from CFFF, and the only direct covariation of CFFF with alpha frequency that we are aware of was reported in hepatic encephalopathy (Butz et al., 2013).

#### 3.1.3 Logical structure and epistemic status

To make the inferential structure explicit: Step 1 established the functional necessity of a stable temporal reference without reference to CFFF; Step 2 introduced CFFF as a candidate marker and evaluated its stability against the prediction. The perceptual clock hypothesis thus generates an independent prediction – that temporal resolution should be stable within and variable between individuals. This prediction is tested against CFFF data rather than derived from it. CFFF stability is an observation consistent with the prediction, not a premise; because stability is also expected under accounts that posit no clock, it supports the hypothesis only weakly. If CFFF were found to be highly unstable within individuals (e.g., intra-individual CV exceeding 15–20%, approaching or surpassing inter-individual variation), this would count against the perceptual clock interpretation even if the theoretical case for a stable clock remained intact.

The relationship between theory and evidence is therefore unidirectional and falsifiable (Figure 3, upper band). This also clarifies the scope of CFFF as a measure: just as a thermometer does not constitute temperature but indexes it with quantifiable reliability and known error, CFFF provides an operationalised readout of temporal-processing capacity. The underlying psychometric properties, including stability, perturbability, and covariance with oscillatory markers, can be independently characterised. The functional necessity of this stable reference is further reinforced by energetic constraints that make faster temporal processing metabolically prohibitive, developed next.

### 3.2 Metabolic and Energetic Constraints

Faster temporal processing requires faster neural dynamics, which involve higher firing rates, shorter membrane time constants, and more rapid synaptic transmission. Each of these factors incurs metabolic cost (Attwell & Laughlin, 2001), and in sensory neurons the cost per bit rises as information rate increases (Laughlin et al., 1998), with synaptic transmission being the largest consumer of neural energy (Sterling & Laughlin, 2015). Transmitting a single bit costs ~$10^4$ ATP molecules at chemical synapses and $10^6$–$10^7$ for spike coding, which is orders of magnitude above the thermodynamic minimum (Laughlin et al., 1998).

Time-averaged control energy for resting brain dynamics correlates with cortical oxygen metabolism (Ceballos et al., 2025), and dynamic functional connectivity reduces energy cost by ~60% relative to static patterns (Deng et al., 2022). These mechanisms indicate active energy optimisation during state transitions. Notably, network-control analyses show that transitions within hierarchical levels (e.g., unimodal–unimodal) are cheaper and more frequent than costly between-level transitions (Ceballos et al., 2025). This pattern is structurally consistent with temporal stability being "built into" the energetics of the primary sensory cortex.

Because faster processing cascades energetically across interconnected systems, individual "optimisation" of temporal resolution may be costly. The brain already operates near metabolic limits – 2% of body mass but 20% of resting metabolism (Raichle & Gusnard, 2002). Furthermore, goal-directed cognition adds only ~5% above resting levels (Jamadar et al., 2025). Kondrakiewicz & Nawrocka (2025) make the complementary argument that brains are metabolically expensive while cognition itself is comparatively cheap, consistent with most energy being committed to fixed infrastructure rather than flexible computation. Even sub-millisecond to few-millisecond variations in conduction delay can shift coupled oscillators from constructive to destructive interference and disrupt inter-regional phase synchrony (Pajevic et al., 2014). As maintaining isochronous conduction across pathways of different length requires costly, activity-dependent myelination, a standardised developmental "clock rate" tuned to species-typical ecological demands may represent an evolutionarily stable solution. These arguments, if correct, account for why a limit exists; they do not by themselves explain why it can be shifted by some classes of experience and not others.

Cross-species comparative evidence is consistent with this account. Using CFFF as the index of maximal temporal information processing, Healy et al. (2013) showed across a wide range of vertebrates that small body size and high mass-specific metabolic rate predict high CFFF. This correlation is consistent with the hypothesis that smaller, faster-metabolising animals can sustain the energetic cost of finer temporal sampling. CFFF spans roughly an order of magnitude across taxa, ranging from low values in large, low-metabolism species to very high values in small, high-metabolism animals. Additionally, some invertebrates such as certain flying insects can resolve flicker well above 200 Hz (reviewed in Mankowska et al., 2021). Such between-species associations do not show that the same relation holds among individuals of one species (Section 6.4).

### 3.3 The Speed–Accuracy Trade-off

Temporal integration embodies a classic speed–accuracy trade-off: longer windows improve signal-to-noise through averaging but sacrifice resolution, while shorter windows improve resolution at the cost of noise sensitivity. The optimal balance is set by the statistics of natural stimuli and the organism's ecological niche (Ho et al., 2012; Mendonça et al., 2020). Natural selection has presumably tuned human temporal integration to balance acuity against noise reduction, and individual learning that pushed resolution beyond this optimum could degrade performance. Even when accuracy is emphasised, observers fail to optimally process sensory signals during speeded decisions (Ho et al., 2012). Furthermore, integration windows that fail to lengthen under motivational manipulation across species point to fundamental neural rather than strategic limits (Mendonça et al., 2020).

Efficient-coding models predict that temporal windows are tuned to natural-scene statistics, with most ecologically relevant events occurring below the flicker-fusion threshold (Ganguli & Simoncelli, 2014; Simoncelli & Olshausen, 2001). Empirically, temporal integration operates across multiple timescales matched to behavioural demands (Deodato & Melcher, 2024; Manea et al., 2024; Spitmaan et al., 2020). The stability of CFFF and the failure of non-specific training to alter fundamental temporal limits (Section 1.2) are consistent with these parameters being constrained by architecture and selection, while the training results in Section 1.2 show that they are not wholly closed to individual learning.

# 4. CFFF as an Individual-Difference Variable: Implications for Cognition

## 4.1 Stable Trait Architecture

The combination of high within-individual stability and substantial between-individual variation (between-person SD of about 5–8 Hz around means of about 47–50 Hz with one LED-based device; about 80% of variance between participants; Section 1.2) makes CFFF a candidate trait marker of neural temporal architecture (Haarlem et al., 2024). Unlike cognitive measures susceptible to practice effects (typically 0.24–0.6 SD improvement; Calamia et al., 2012), CFFF requires only brief practice before reaching a stable plateau (Parkin et al., 1997), although a continued rise across sessions has been reported in women (Haarlem et al., 2024). The plateau indicates that the familiarisation effect saturates below the threshold of detectable change, which allows it to be distinguished from training-induced gains; it does not by itself establish that the measure is untrainable.

These trait properties coexist with predictable state-dependent and demographic modulation, which the framework accommodates rather than treats as anomalous. Sex and hormonal status may influence CFFF, though the evidence is inconsistent: Haarlem et al. (2024) found no significant mean difference between men and women, with only greater test–retest variability observed in women, while reports of menstrual-cycle effects diverge in direction as some studies find reduced visual sensitivity premenstrually (Ward et al., 1978) and others find reduced performance around ovulation on a different visual task (Scher et al., 1985), a discrepancy the latter authors attribute to task-dependent differences in the visual sub-processes being measured. Rather than undermining the stability thesis, such effects – in whichever direction they run – exemplify the state-dependent modulation of trait expression described in Section 2.1.2. They are expected to shift current performance without shifting the underlying limit; whether between-individual rankings are preserved across such states has not been tested. Circadian and seasonal variation behave similarly, representing predictable patterns superimposed on stable individual differences (Łuczak & Sobolewski, 2005; Walsh & Misiak, 1966).

CFFF declines with age (Lachenmayr et al., 1994; Muth et al., 2026; Wolf & Schraffa, 1964). Whether the decline accelerates in later life is unclear: flicker perimetry in 130 people aged 9 to 86 years showed a linear loss of about 4% per decade with no acceleration (Lachenmayr et al., 1994), whereas another sample showed the largest reductions in the 60- and 70-year age groups (Muth et al., 2026). Neurophysiologically, ageing selectively increases the latency of magnocellular-dominated visual evoked potentials. This delay is steepest in second-order nonlinear components reflecting cortical processing efficiency (~6 ms/decade), while first-order responses show no significant latency shift (Brown et al., 2019). These findings suggest that part of the age-related decline arises in cortical temporal integration rather than in photoreceptor function. Because practice effects on CFFF are small and saturate early, they are less likely than for cognitive tests to obscure longitudinal decline (Salthouse, 2010), although the rise of about 5% in some untrained controls (Section 1.2) shows that they are not negligible.

Evidence bearing directly on the trait-marker proposal is, however, mixed, including in data from two recent studies. In a sample of 36 healthy adults tested across three sessions with the Simon, flanker, Corsi block-tapping and digit span tasks, flicker measures were stable across sessions, as the present account predicts, but their associations with cognitive performance were weak and sex-dependent: no significant correlations were observed among men, whereas among women flicker frequency was related to backward digit span ($r = 0.460$) and to the flanker effect ($r = -0.360$) (Mankowska et al., 2025). A subsequent study in which 40 healthy adults were tested at 4 ATA under three breathing gases found that CFFF and reaction time both rose modestly during exposure but diverged after decompression, indicating that the two measures do not track a common underlying process in any simple way (Mankowska et al., 2026). The cognitive batteries in both studies were general rather than selected to isolate temporal-processing demands, which limits what can be concluded from the weak associations; the findings cannot, however, be set aside. As matters stand, the stability component of the trait-marker claim is supported whereas its predictive component is not, and the hypotheses developed in Sections 4.2 and 4.3 should be read against that asymmetry rather than as extensions of an established finding.

### 4.2 CFFF and Working Memory

The hypotheses that follow are therefore advanced in the presence of a first, largely negative test, and are framed to explain why that test may have been insensitive rather than to assume that it was. Theoretical and empirical work suggests links between temporal resolution and working memory, which we frame as hypotheses rather than established causal relations. Pan (2020) showed that active maintenance of information in visual working memory reduces temporal resolution, impairing concurrent detection of brief gaps. This bidirectional interaction is consistent with shared neural resources or constraints. Concretely, loading working memory with a concurrent task should depress measured CFFF relative to an unloaded baseline; this is directly testable. Furthermore, Li et al. (2021) showed that working-memory representations are encoded as probabilistic distributions whose width reflects uncertainty, with larger decoded uncertainty predicting greater behavioural error.

Together these findings motivate the hypothesis that temporal-processing capacity may index a bottleneck on higher cognition. Oscillatory mechanisms plausibly mediate this link. Specifically, theta oscillations coordinate temporal sequences across hippocampal–cortical networks (Etter et al., 2023), and theta–gamma coupling organises multiple items through phase-coded representations (Lisman & Jensen, 2013; Pirazzini & Ursino, 2024). If maintenance requires rapid temporal updating through such mechanisms, the maximum temporal resolution of the system, as indexed by CFFF, may set an upper bound on how many items can be independently maintained. This would amount to a restatement of classic capacity limits (Cowan, 2001; Miller, 1956) in temporal terms. We emphasise that the directionality of these relationships (whether CFFF constrains capacity, co-varies with it, or both) remains to be established by manipulation studies (Section 6.1).

### 4.3 Constraint Propagation Through Hierarchical Architectures

We propose that constraints at lower levels of the processing hierarchy propagate upward, limiting what higher-level learning can achieve. We use the term *constraint propagation* in a specific sense that distinguishes it from related ideas.

Unlike a simple bottleneck model, in which a single stage caps throughput, constraint propagation concerns how a fixed parameter at one level (temporal resolution) reshapes the representational format available at all higher levels, not merely their rate. Unlike cascade-failure models, it does not require degradation or breakdown – it operates in the normally functioning system. Finally, unlike resource-competition models, which describe dynamic, reallocable trade-offs between concurrent processes, constraint propagation concerns a boundary condition that ordinary learning does not renegotiate. The term propagation is suitable because the consequences of the fixed temporal grain are inherited, not re-derived, at each successive level.

This concept can be illustrated by two examples. First, attentional sampling operates at frequencies constrained by early visual processing. For instance, the attentional blink (~200–500 ms) may reflect temporal-integration constraints related to those indexed by CFFF (Willems & Martens, 2016; Zivony & Lamy, 2022). Based on this, we predict that individuals with higher CFFF show shorter attentional blinks, though this correlational test would not yet establish causation. Second, the temporal resolution of perceptual decisions is bounded by sensory sampling: if CFFF reflects the maximum temporal bandwidth of early vision, even high-level decisions cannot access finer temporal scales. This is consistent with the temporal specificity of perceptual learning, such as interval-specific gains in auditory interval discrimination that do not transfer to neighbouring intervals (Karmarkar & Buonomano, 2003) or between single-interval and two-interval discrimination tasks trained on the same interval (Xu et al., 2021). This pattern contrasts with the broader transfer seen in spatial perceptual learning (Kahalani-Hodedany et al., 2024), which would be consistent with temporal resolution being less readily modified than spatial resolution.

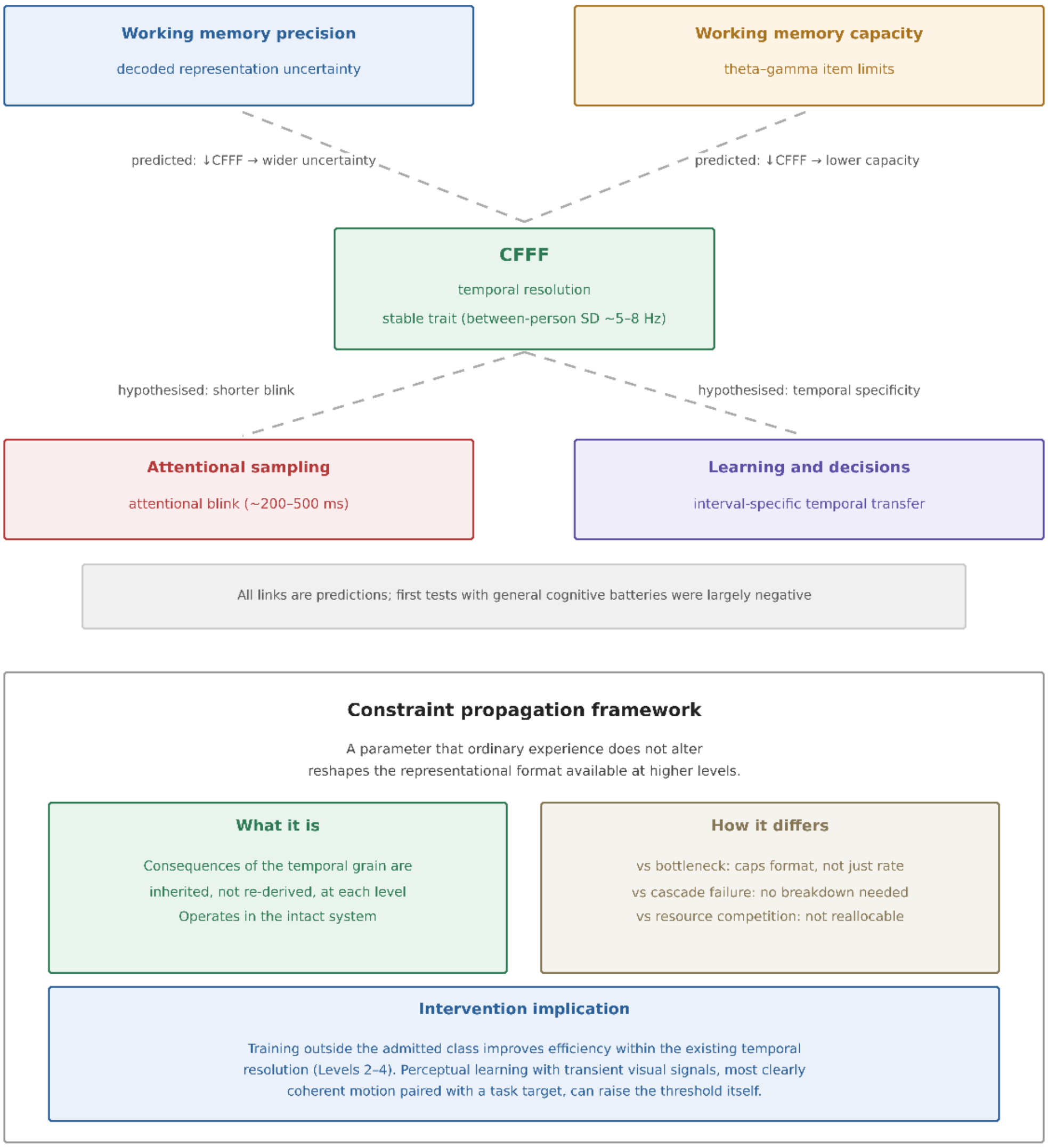

Figure 4. CFFF as a stable trait marker: hypothesised constraint propagation through cognitive architecture. A stable CFFF trait is linked by predicted (dashed), not established, relationships to working-memory precision and capacity, attentional sampling, and perceptual learning and decision processes. The framework (lower panel) defines constraint propagation as the inheritance of a low-level temporal parameter by higher levels and distinguishes it from bottleneck, cascade-failure and resource-competition models. Training outside the admitted class of experience is predicted to improve efficiency within the existing temporal resolution (Levels 2-4) without raising the threshold itself, whereas training within that class can raise it (Section 1.2). All depicted links are predictions awaiting empirical test; first tests using general cognitive batteries were largely negative (Section 4.1).

## 5. Implications for Theoretical Frameworks in Cognitive Neuroscience

A full integration of CFFF with each major theoretical framework is beyond the scope of this review. Instead, for each we draw out a single, focused implication of treating temporal resolution as a parameter whose modification is restricted.

### 5.1 Predictive Coding and the Free-Energy Principle

Predictive coding assumes that the generative model can be optimised through learning (Friston & Kiebel, 2009; Rao & Ballard, 1999). Properties whose modification is restricted, such as CFFF, suggest a two-tier architecture: learnable parameters (priors, precision weights, synaptic prediction weights; Feldman & Friston, 2010; Tschantz et al., 2023) operate on top of fixed or only narrowly modifiable parameters (temporal kernels, sampling rates indexed by CFFF, sensory noise floors). Intrinsic neural timescales, short in sensory cortex (~50–100 ms) and longer in higher-order regions, have been proposed to reflect structural constraints (Ibanez & Northoff, 2024), although training can alter timescales in prefrontal cortex (Trepka et al., 2024). The focused implication is that models attempting to optimise all parameters may misrepresent the limits on perceptual learning (Jiang & Rao, 2024). A model that respects fixed temporal infrastructure may better account for individual differences and for cases such as atypical perceptual learning in autism despite intact basic inference (Fazioli et al., 2025).

### 5.2 Bayesian Brain and Optimal Inference

Under the Bayesian-brain hypothesis, perception combines likelihood with prior knowledge (Kersten et al., 2004; Knill & Richards, 1996). CFFF sets an upper limit on the temporal resolution of the likelihood: an ideal observer with perfect inference but limited temporal resolution will still err on tasks requiring temporal precision beyond that limit. The focused, testable implication is that individuals with lower CFFF should rely more heavily on priors to compensate for temporally noisier evidence, producing stronger central-tendency biases – a pattern already documented when sensory precision is low (Cicchini et al., 2012; Jazayeri & Shadlen, 2010; Turgeon et al., 2016). This predicts a correlation between CFFF and the strength of prior reliance in temporal judgment, which remains to be tested directly.

### 5.3 Neural Darwinism and Selectionist Theories

Selectionist theories hold that learning selects among pre-existing circuits rather than constructing new ones (Edelman, 1987; Sporns & Edelman, 1993). If the restriction described here is genuine, not all properties are equally available for selection in adulthood. The focused implication is a hybrid model in which selection acts on connectivity (plastic) within fixed biophysical boundaries. Critical periods establish these boundaries in a hierarchical cascade (Nakamura et al., 2020), after which membrane properties, synaptic time constants, and intrinsic oscillatory frequencies define the temporal bandwidth for subsequent learning (Buzi et al., 2024), which would be consistent with CFFF remaining stable despite lifelong synaptic plasticity.

## 6. Future Research Directions

### 6.1 CFFF and Working-Memory Uncertainty

A priority is to test directly whether CFFF predicts the precision of working-memory representations. Two recent studies have made a first pass at this question using general neuropsychological batteries and found weak, sex-dependent or divergent associations (Mankowska et al., 2025, 2026). The designs proposed here differ in two respects that may prove decisive: they employ measures selected for their temporal-processing demands rather than general cognitive tests, and they manipulate temporal load rather than only correlating across individuals. If associations remain absent under those conditions, the constraint-propagation account of Section 4.3 loses its principal motivation. Using neural decoding (Li et al., 2021), distribution width during maintenance could be related to individual CFFF – if CFFF reflects fundamental temporal-resolution constraints, lower CFFF should predict wider uncertainty distributions for the same memoranda. A complementary manipulation tests directionality: if temporal resources are shared between CFFF-level processing and maintenance, working-memory load should depress CFFF (extending Pan, 2020) while temporal demands should reduce memory precision, with bidirectional interference indicating shared resource pools (Fortin & Rousseau, 1998). We frame four specific, falsifiable predictions: lower CFFF should be associated with (1) wider decoded uncertainty distributions (Broadway & Engle, 2011); (2) poorer metacognitive calibration (Conte et al., 2023; Komori, 2016); (3) reduced effective capacity as temporal interference accumulates with set size (Zhang & Luck, 2008); and (4) greater susceptibility to temporal crowding over short and long intervals (Hochmitz et al., 2024). Each prediction specifies a measurable outcome whose absence would constrain the constraint-propagation account.

### 6.2 Interventions Targeting Temporal Processing

Apart from the restricted class of perceptual learning described in Section 1.2, interventions may alter how efficiently available resolution is used rather than the resolution itself. Action-video-game training improves temporal attention and reduces the attentional blink (Gan et al., 2020; Green & Bavelier, 2003), with some controlled training studies reporting benefits (Oei & Patterson, 2013), although meta-analyses disagree on the size and generality of such effects (Bediou et al., 2018; Sala et al., 2018), and benefits extending to amblyopia (Li et al., 2011) and dyslexia under parietal neuromodulation (Bertoni et al., 2024). Where present, these effects more plausibly reflect optimised attentional allocation rather than changed fundamental resolution, supporting the hierarchical model. Pharmacologically, temporal processing is modulated by several neurotransmitter systems, all in a state-dependent, reversible manner consistent with a stable underlying limit. Dopamine is best characterised – D2 activity in the basal ganglia supports millisecond timing (Meck, 1996; Rammsayer, 1997), Parkinson's disease produces timing deficits (Harrington et al., 2011; Jones et al., 2008), and methylphenidate improves them with baseline D1/D2 ratios predicting response (Manza et al., 2025; Rubia et al., 2009) – but cholinergic, serotonergic, and adenosinergic systems also modulate cortical arousal and flicker sensitivity and are particularly relevant to ageing and neurodegeneration (Hindmarch, 1982). The dose-dependent, reversible nature of these effects, and the persistence of deficits after chronic dopaminergic dysfunction (Harrington et al., 2011), are consistent with pharmacological agents modulating the efficiency of existing temporal mechanisms rather than the limits of achievable resolution.

### 6.3 Testing the account: consolidation, class restriction and the CFFF–alpha link

The most direct extension of this framework is to test its own predictions (Section 1.2), of which the consolidation prediction is currently the least constrained by data. The decisive study trains participants with a paradigm from the provisionally admitted class (Seitz et al., 2006; Zhou et al., 2016) and then measures retention at 1 week, 1 month, and 6 months against an active control matched for engagement but not for temporal content, with the threshold measured with both the small-field densitometer stimulus and a standard LED-based procedure so that the construct-generality prediction is tested in the same sample. Consolidated gains that persist beyond one week would support the consolidation prediction and confirm that targeted training produces perceptual learning rather than transient arousal. Rapid decay toward baseline, by contrast, would reclassify the effect as state modulation and would require revising how the training literature is interpreted. The same design also sharpens the class-restriction prediction, since the control arm provides the null benchmark against which any genuine temporal-stream effect must be judged.

A second study tests the link on which the perceptual clock rationale depends. If CFFF indexes a stable oscillatory pacemaker (Section 3.1), individual alpha frequency (IAF) should covary with CFFF across individuals, and – more stringently – causally driving alpha above or below each participant's IAF with transcranial alternating current stimulation should shift CFFF predictably, mirroring the established effect of such stimulation on the multisensory temporal binding window (Cecere et al., 2015; Samaha & Postle, 2015). A predicted shift would indicate that CFFF depends on the clock rate rather than merely correlating with it. A null causal effect, despite a reliable IAF-CFFF correlation, would indicate that the two share an upstream constraint without a direct generative link, narrowing the mechanistic claim. Inhibitory stimulation of the left inferior parietal lobule, already shown to alter the flicker threshold (Nardella et al., 2014), offers a complementary causal probe of the network's read-out node.

### 6.4 Extending the framework: neuroimaging, cross-species, and clinical directions

Three further directions follow from claims the review makes but does not itself test. First, the dissociation argued in Section 2.2 is ultimately a neuroimaging hypothesis: longitudinal fMRI or EEG before and after spatial-learning and temporal-learning training should show that CFFF stability tracks the stability of feedforward thalamocortical signatures, whereas trained spatial gains track plasticity in horizontal and feedback connectivity. Co-variation of behavioural change with feedforward rather than recurrent markers would support the architectural account; change localised to feedforward pathways alongside stable CFFF would falsify it. Second, the metabolic argument (Section 3.2; Healy et al., 2013) generates an untested within-species prediction: if temporal resolution is bounded by whole-organism energy budget, individual differences in human CFFF should correlate with mass-specific metabolic and body-size markers, providing a bridge between the cross-species pattern and the individual-difference framework of Section 4. Third, the scattered clinical evidence motivates a biomarker question: if CFFF is a stable individual trait, an individual's deviation from their own baseline may be more diagnostically sensitive than comparison to population norms in conditions marked by cortical oscillatory slowing, such as hepatic encephalopathy (Butz et al., 2013), multiple sclerosis, and disorders of consciousness. A study comparing intra-individual baseline-referenced CFFF against population-referenced scoring would test whether trait stability translates into clinical utility.

### 6.5 Where the gate lies

Finally, the framework should be pressed on its own limits. Because targeted training, and flicker training in amblyopic eyes starting below typical values, do modify CFFF (Supplementary Table S1), the open question is not whether CFFF can change but what the gate admits: which paradigms, training intensities and populations yield durable change, and which do not. Systematically mapping these conditions converts the apparent exception into a research programme and specifies the framework's domain of validity, identifying which conditions at Level 1 (Figure 1) separate stable expression from consolidated modification.

## 7. Conclusion

Critical flicker fusion frequency is best described neither as fixed nor as freely modifiable, but as an experience-restricted property: it is stable within individuals, little affected by repeated testing and unaffected by training without temporal content regardless of dose, and raised only by a restricted class of perceptual learning, most clearly by task-irrelevant coherent motion temporally paired with task targets. The contribution of this review is to state that claim in a form that can fail. It specifies three conditions for an experience-restricted constraint and five predictions, of which the most consequential is that gains measured with the small-field densitometer stimulus replicate with standard LED-based measurement. The evidence that meets these conditions is thin: training samples of three to ten participants, retention data from three people, a single measurement procedure for normally sighted adults, and a minority of untrained controls who changed as much as trained ones. Three further elements are interpretive rather than established. The attribution of the restriction to critical-period-stabilised thalamocortical circuitry is a hypothesis without direct human evidence. The perceptual clock, metabolic and speed–accuracy arguments are rationales for why such a restriction might be adaptive, not evidence that it exists. The proposed links between CFFF and working-memory precision, metacognition and capacity are predictions, and the first tests of them were largely negative. Adequately powered training studies with active controls, pre-specified retention intervals and both measurement procedures are the necessary next step; they would show whether CFFF is a useful model of how the class of experience, rather than its amount, gates modification of a basic perceptual property.

**Declaration of Competing Interest:** The authors declare that they have no known competing financial interests or personal relationships that could have appeared to influence the work reported in this paper.

**Funding:** This research did not receive any specific grant from funding agencies in the public, commercial, or not-for-profit sectors.

**Open Practices Statement:** This review reports no new data, materials or analysis code; all data discussed are available in the cited published sources. The review was not preregistered.

## References

Astorga, G., Chen, M., Yan, Y., Altavini, T. S., Jiang, C. S., Li, W., & Gilbert, C. (2022). Adaptive processing and perceptual learning in visual cortical areas V1 and V4. *Proceedings of the National Academy of Sciences*, *119*(42), e2213080119. https://doi.org/10.1073/pnas.2213080119

Attwell, D., & Laughlin, S. B. (2001). An Energy Budget for Signaling in the Grey Matter of the Brain. *Journal of Cerebral Blood Flow & Metabolism*, *21*(10), 1133–1145. https://doi.org/10.1097/00004647-200110000-00001

Balestra, C., Lafere, P., & Germonpre, P. (2012). Persistence of critical flicker fusion frequency impairment after a 33 mfw SCUBA dive: Evidence of prolonged nitrogen narcosis? *European Journal of Applied Physiology*, *112*(12), 4063–4068. https://doi.org/10.1007/s00421-012-2391-z

Balestra, C., Machado, M.-L. L., Theunissen, S., Balestra, A., Cialoni, D., Clot, C., Besnard, S., Kammacher, L., Delzenne, J., Germonpré, P., & Lafère, P. (2018). Critical flicker fusion frequency: A marker of cerebral arousal during modified gravitational conditions related to parabolic flights. *Frontiers in Physiology*, *9*, 1403. https://doi.org/10.3389/fphys.2018.01403

Baror, S., Baumgarten, T., & He, B. J. (2024). Neural mechanisms determining the duration of task-free, self-paced visual perception. *Journal of Cognitive Neuroscience*, *36*(5), 756–775. https://doi.org/10.1162/jocn_a_02131

Bavelier, D., Levi, D. M., Li, R. W., Dan, Y., & Hensch, T. K. (2010). Removing brakes on adult brain plasticity: From molecular to behavioral interventions. *The Journal of Neuroscience*, *30*(45), 14964–14971. https://doi.org/10.1523/JNEUROSCI.4812-10.2010

Bediou, B., Adams, D. M., Mayer, R. E., Tipton, E., Green, C. S., & Bavelier, D. (2018). Meta-analysis of action video game impact on perceptual, attentional, and cognitive skills. *Psychological Bulletin, 144*(1), 77–110. https://doi.org/10.1037/bul0000130

Bernardi, L., Costa, V. P., & Oto Shiroma, L. (2007). Flicker perimetry in healthy subjects: Influence of age and gender, learning effect and short-term fluctuation. *Arquivos Brasileiros de Oftalmologia*, *70*(1), 91–99. https://doi.org/10.1590/S0004-27492007000100017

Bertoni, S., Franceschini, S., Mancarella, M., Puccio, G., Ronconi, L., Marsicano, G., Gori, S., Campana, G., & Facoetti, A. (2024). Action video games and posterior parietal cortex neuromodulation enhance both attention and reading in adults with developmental dyslexia. *Cerebral Cortex*, *34*(4), bhae152. https://doi.org/10.1093/cercor/bhae152

Bharmauria, V., Ouelhazi, A., Lussiez, R., & Molotchnikoff, S. (2022). Adaptation-induced plasticity in the sensory cortex. *Journal of Neurophysiology*, *128*(4), 946–962. https://doi.org/10.1152/jn.00114.2022

Bovier, E. R., Renzi, L. M., & Hammond, B. R. (2014). A double-blind, placebo-controlled study on the effects of lutein and zeaxanthin on neural processing speed and efficiency. *PLOS ONE, 9*(9), e108178. https://doi.org/10.1371/journal.pone.0108178

Broadway, J. M., & Engle, R. W. (2011). Individual Differences in Working Memory Capacity and Temporal Discrimination. *PLOS ONE*, *6*(10), e25422. https://doi.org/10.1371/journal.pone.0025422

Brock, J., Brown, C. C., Boucher, J., & Rippon, G. (2002). The temporal binding deficit hypothesis of autism. *Development and Psychopathology*, *14*(2), 209–224. https://doi.org/10.1017/S0954579402002018

Brown, A., Corner, M., Crewther, D., & Crewther, S. (2019). Age Related Decline in Cortical Multifocal Flash VEP: Latency Increases Shown to Be Predominately Magnocellular. *Frontiers in Aging Neuroscience*, *10*, 430. https://doi.org/10.3389/fnagi.2018.00430

Brown, A., Corner, M., Crewther, D. P., & Crewther, S. G. (2018). Human Flicker Fusion Correlates With Physiological Measures of Magnocellular Neural Efficiency. *Frontiers in Human Neuroscience*, *12, Article 176.* https://doi.org/10.3389/fnhum.2018.00176

Burns, M. E., & Baylor, D. A. (2001). Activation, Deactivation, and Adaptation in Vertebrate Photoreceptor Cells. *Annual Review of Neuroscience*, *24,* 779–805. https://doi.org/10.1146/annurev.neuro.24.1.779

Butz, M., May, E. S., Häussinger, D., & Schnitzler, A. (2013). The slowed brain: Cortical oscillatory activity in hepatic encephalopathy. *Archives of Biochemistry and Biophysics, 536*(2), 197–203. https://doi.org/10.1016/j.abb.2013.04.004

Buzi, G., Eustache, F., Droit-Volet, S., Desaunay, P., & Hinault, T. (2024). Towards a neurodevelopmental cognitive perspective of temporal processing. *Communications Biology*, *7*(1), 987. https://doi.org/10.1038/s42003-024-06641-4

Calamia, M., Markon, K., & Tranel, D. (2012). Scoring Higher the Second Time Around: Meta-Analyses of Practice Effects in Neuropsychological Assessment. *The Clinical Neuropsychologist*, *26*(4), 543–570. https://doi.org/10.1080/13854046.2012.680913

Ceballos, E. G., Luppi, A. I., Castrillon, G., Saggar, M., Misic, B., & Riedl, V. (2025). The control costs of human brain dynamics. *Network Neuroscience*, *9*(1), 77–99. https://doi.org/10.1162/netn_a_00425

Cecere, R., Rees, G., & Romei, V. (2015). Individual differences in alpha frequency drive crossmodal illusory perception. *Current Biology, 25*(2), 231–235. https://doi.org/10.1016/j.cub.2014.11.034

Censor, N., & Sagi, D. (2009). Global resistance to local perceptual adaptation in texture discrimination. *Vision Research, 49*(21), 2550–2556. https://doi.org/10.1016/j.visres.2009.03.018

Chaudhuri, R., Knoblauch, K., Gariel, M.-A., Kennedy, H., & Wang, X.-J. (2015). A Large-Scale Circuit Mechanism for Hierarchical Dynamical Processing in the Primate Cortex. *Neuron*, *88*(2), 419–431. https://doi.org/10.1016/j.neuron.2015.09.008

Chen, J. L., Lin, W. C., Cha, J. W., So, P. T., Kubota, Y., & Nedivi, E. (2011). Structural basis for the role of inhibition in facilitating adult brain plasticity. *Nature Neuroscience*, *14*(5), 587–594. https://doi.org/10.1038/nn.2799

Chen, N., Sugihara, H., & Sur, M. (2015). An acetylcholine-activated microcircuit drives temporal dynamics of cortical activity. *Nature Neuroscience*, *18*(6), 892–902. https://doi.org/10.1038/nn.4002

Chung, S., Jeong, J.-H., Ko, S., Yu, X., Kim, Y.-H., Isaac, J. T. R., & Koretsky, A. P. (2017). Peripheral Sensory Deprivation Restores Critical-Period-like Plasticity to Adult Somatosensory Thalamocortical Inputs. *Cell Reports*, *19*(13), 2707–2717. https://doi.org/10.1016/j.celrep.2017.06.018

Cicchini, G. M., Arrighi, R., Cecchetti, L., Giusti, M., & Burr, D. C. (2012). Optimal Encoding of Interval Timing in Expert Percussionists. *The Journal of Neuroscience*, *32*(3), 1056–1060. https://doi.org/10.1523/JNEUROSCI.3411-11.2012

Cisneros-Franco, J. M., & de Villers-Sidani, É. (2019). Reactivation of critical period plasticity in adult auditory cortex through chemogenetic silencing of parvalbumin-positive interneurons. *Proceedings of the National Academy of Sciences*, *116*(52), 26329–26331. https://doi.org/10.1073/pnas.1913227117

Conci, M., Busch, N., Rozek, R. P., & Müller, H. J. (2023). Learning-Induced Plasticity Enhances the Capacity of Visual Working Memory. *Psychological Science*, *34*(10), 1087–1100. https://doi.org/10.1177/09567976231192241

Conte, N., Fairfield, B., Padulo, C., & Pelegrina, S. (2023). Metacognition in working memory: Confidence judgments during an n-back task. *Consciousness and Cognition*, *111*, 103522. https://doi.org/10.1016/j.concog.2023.103522

Cowan, N. (2001). The magical number 4 in short-term memory: A reconsideration of mental storage capacity. *Behavioral and Brain Sciences*, *24*(1), 87–114. https://doi.org/10.1017/s0140525x01003922

Daw, N. W., Fox, K., Sato, H., & Czepita, D. (1992). Critical period for monocular deprivation in the cat visual cortex. *Journal of Neurophysiology*, *67*(1), 197–202. https://doi.org/10.1152/jn.1992.67.1.197

Debanne, D., Campanac, E., Bialowas, A., Carlier, E., & Alcaraz, G. (2011). Axon physiology. *Physiological Reviews*, *91*(2), 555–602. https://doi.org/10.1152/physrev.00048.2009

Deng, S., Li, J., Thomas Yeo, B. T., & Gu, S. (2022). Control theory illustrates the energy efficiency in the dynamic reconfiguration of functional connectivity. *Communications Biology*, *5*(1), 295. https://doi.org/10.1038/s42003-022-03196-0

Deodato, M., & Melcher, D. (2024). Continuous temporal integration in the human visual system. *Journal of Vision*, *24*(13), 5. https://doi.org/10.1167/jov.24.13.5

Edelman, G. M. (1987). *Neural Darwinism: The theory of neuronal group selection*. Basic Books. http://archive.org/details/neuraldarwinismt0000edel

Eisen-Enosh, A., Farah, N., Burgansky-Eliash, Z., Polat, U., & Mandel, Y. (2017). Evaluation of Critical Flicker-Fusion Frequency Measurement Methods for the Investigation of Visual Temporal Resolution. *Scientific Reports*, *7*(1), 15621. https://doi.org/10.1038/s41598-017-15034-z

Eisen-Enosh, A., Farah, N., Polat, U., & Mandel, Y. (2023). Perceptual learning based on a temporal stimulus enhances visual function in adult amblyopic subjects. *Scientific Reports*, *13*(1), 7643. https://doi.org/10.1038/s41598-023-34421-3

Espinosa, J. S., & Stryker, M. P. (2012). Development and Plasticity of the Primary Visual Cortex. *Neuron*, *75*(2), 230–249. https://doi.org/10.1016/j.neuron.2012.06.009

Etter, G., Carmichael, J. E., & Williams, S. (2023). Linking temporal coordination of hippocampal activity to memory function. *Frontiers in Cellular Neuroscience*, *17*. https://doi.org/10.3389/fncel.2023.1233849

Fazioli, L., Hadad, B.-S., Denison, R. N., & Yashar, A. (2025). Suboptimal but intact integration of Bayesian components during perceptual decision-making in autism. *Molecular Autism*, *16*(1), 2. https://doi.org/10.1186/s13229-025-00639-3

Feldman, H., & Friston, K. (2010). Attention, Uncertainty, and Free-Energy. *Frontiers in Human Neuroscience*, *4*. https://doi.org/10.3389/fnhum.2010.00215

Ferguson, M. A., Henshaw, H., Clark, D. P. A., & Moore, D. R. (2014). Benefits of Phoneme Discrimination Training in a Randomized Controlled Trial of 50- to 74-Year-Olds With Mild Hearing Loss. *Ear and Hearing*, *35*(4), e110–e121. https://doi.org/10.1097/AUD.0000000000000020

Fornaciai, M., & Park, J. (2018). Attractive Serial Dependence in the Absence of an Explicit Task. *Psychological Science*, *29*(3), 437–446. https://doi.org/10.1177/0956797617737385

Fortin, C., & Rousseau, R. (1998). Interference from short-term memory processing on encoding and reproducing brief durations. *Psychological Research*, *61*(4), 269–276. https://doi.org/10.1007/s004260050031

Foss-Feig, J. H., Kwakye, L. D., Cascio, C. J., Burnette, C. P., Kadivar, H., Stone, W. L., & Wallace, M. T. (2010). An extended multisensory temporal binding window in autism spectrum disorders. *Experimental Brain Research*, *203*(2), 381–389. https://doi.org/10.1007/s00221-010-2240-4

Freschl, J., Azizi, L. A., Balboa, L., Kaldy, Z., & Blaser, E. (2022). The development of peak alpha frequency from infancy to adolescence and its role in visual temporal processing: A meta-analysis. *Developmental Cognitive Neuroscience*, *57*, 101146. https://doi.org/10.1016/j.dcn.2022.101146

Freschl, J., Melcher, D., Kaldy, Z., & Blaser, E. (2019). Visual temporal integration windows are adult-like in 5- to 7-year-old children. *Journal of Vision*, *19*(7), 5. https://doi.org/10.1167/19.7.5

Friston, K., & Kiebel, S. (2009). Predictive coding under the free-energy principle. *Philosophical Transactions of the Royal Society B: Biological Sciences*, *364*(1521), 1211–1221. https://doi.org/10.1098/rstb.2008.0300

Froemke, R. C. (2015). Plasticity of Cortical Excitatory-Inhibitory Balance. *Annual Review of Neuroscience*, *38,* 195–219. https://doi.org/10.1146/annurev-neuro-071714-034002

Fu, Y.-X., Djupsund, K., Gao, H., Hayden, B., Shen, K., & Dan, Y. (2002). Temporal specificity in the cortical plasticity of visual space representation. *Science, 296*(5575), 1999–2003. https://doi.org/10.1126/science.1070521

Fujisaki, W., Shimojo, S., Kashino, M., & Nishida, S. (2004). Recalibration of audiovisual simultaneity. *Nature Neuroscience*, *7*(7), 773–778. https://doi.org/10.1038/nn1268

Gan, X., Yao, Y., Liu, H., Zong, X., Cui, R., Qiu, N., Xie, J., Jiang, D., Ying, S., Tang, X., Dong, L., Gong, D., Ma, W., & Liu, T. (2020). Action Real-Time Strategy Gaming Experience Related to Increased Attentional Resources: An Attentional Blink Study. *Frontiers in Human Neuroscience*, *14*. https://doi.org/10.3389/fnhum.2020.00101

Ganguli, D., & Simoncelli, E. P. (2014). Efficient sensory encoding and Bayesian inference with heterogeneous neural populations. *Neural Computation*, *26*(10), 2103–2134. https://doi.org/10.1162/NECO_a_00638

Gervain, J., Vines, B. W., Chen, L. M., Seo, R. J., Hensch, T. K., Werker, J. F., & Young, A. H. (2013). Valproate reopens critical-period learning of absolute pitch. *Frontiers in Systems Neuroscience*, *7*. https://doi.org/10.3389/fnsys.2013.00102

Gilbert, C. D., & Li, W. (2012). Adult visual cortical plasticity. *Neuron*, *75*(2), 250–264. https://doi.org/10.1016/j.neuron.2012.06.030

Gilbert, C. D., Li, W., & Piech, V. (2009). Perceptual learning and adult cortical plasticity. *The Journal of Physiology*, *587*(Pt 12), 2743–2751. https://doi.org/10.1113/jphysiol.2009.171488

Gordon, J. A., & Stryker, M. P. (1996). Experience-dependent plasticity of binocular responses in the primary visual cortex of the mouse. *The Journal of Neuroscience, 16*(10), 3274–3286. https://doi.org/10.1523/JNEUROSCI.16-10-03274.1996

Görtelmeyer, R., & Wiemann, H. (1982). Retest Reliability and Construct Validity of Critical Flicker Fusion Frequency. *Pharmacopsychiatry*, *15*, 24–28. https://doi.org/10.1055/s-2007-1019545

Green, C. S., & Bavelier, D. (2003). Action video game modifies visual selective attention. *Nature*, *423*(6939), 534–537. https://doi.org/10.1038/nature01647

Haarlem, C., O'Connell, R., Mitchell, K., & Jackson, A. (2024). The speed of sight: Individual variation in critical flicker fusion thresholds. *PLOS ONE*, *19(4), e0298007.* https://doi.org/10.1371/journal.pone.0298007

Harrington, D. L., Castillo, G. N., Greenberg, P. A., Song, D. D., Lessig, S., Lee, R. R., & Rao, S. M. (2011). Neurobehavioral mechanisms of temporal processing deficits in Parkinson's disease. *PLOS ONE, 6*(2), e17461. https://doi.org/10.1371/journal.pone.0017461

Healy, K., McNally, L., Ruxton, G. D., Cooper, N., & Jackson, A. L. (2013). Metabolic rate and body size are linked with perception of temporal information. *Animal Behaviour*, *86*(4), 685–696. https://doi.org/10.1016/j.anbehav.2013.06.018

Hensch, T. K. (2005). Critical period plasticity in local cortical circuits. *Nature Reviews Neuroscience*, *6*(11), 877–888. https://doi.org/10.1038/nrn1787

Hensch, T. K., Fagiolini, M., Mataga, N., Stryker, M. P., Baekkeskov, S., & Kash, S. F. (1998). Local GABA circuit control of experience-dependent plasticity in developing visual cortex. *Science, 282*(5393), 1504–1508. https://doi.org/10.1126/science.282.5393.1504

Hindmarch, I. (1982). Critical Flicker Fusion Frequency (CFF): The Effects of Psychotropic Compounds. *Pharmacopsychiatry*, *15*, 44–48. https://doi.org/10.1055/s-2007-1019549

Ho, T., Brown, S., van Maanen, L., Forstmann, B. U., Wagenmakers, E.-J., & Serences, J. T. (2012). The optimality of sensory processing during the speed-accuracy tradeoff. *The Journal of Neuroscience, 32*(23), 7992–8003. https://doi.org/10.1523/JNEUROSCI.0340-12.2012

Hochmitz, I., Abu-Akel, A., & Yeshurun, Y. (2024). Interference across time: Dissociating short from long temporal interference. *Frontiers in Psychology*, *15*. https://doi.org/10.3389/fpsyg.2024.1393065

Hooks, B. M., & Chen, C. (2006). Distinct roles for spontaneous and visual activity in remodeling of the retinogeniculate synapse. *Neuron*, *52*(2), 281–291. https://doi.org/10.1016/j.neuron.2006.07.007

Hooks, B. M., & Chen, C. (2007). Critical periods in the visual system: Changing views for a model of experience-dependent plasticity. *Neuron*, *56*(2), 312–326. https://doi.org/10.1016/j.neuron.2007.10.003

Hua, L., Gao, F., Leong, C., & Yuan, Z. (2023). Neural decoding dissociates perceptual grouping between proximity and similarity in visual perception. *Cerebral Cortex*, *33*(7), 3803–3815. https://doi.org/10.1093/cercor/bhac308

Ibanez, A., & Northoff, G. (2024). Intrinsic timescales and predictive allostatic interoception in brain health and disease. *Neuroscience and Biobehavioral Reviews*, *157*, 105510. https://doi.org/10.1016/j.neubiorev.2023.105510

Jamadar, S. D., Behler, A., Deery, H., & Breakspear, M. (2025). The metabolic costs of cognition. *Trends in Cognitive Sciences*, *29*(6), 541–555. https://doi.org/10.1016/j.tics.2024.11.010

Jazayeri, M., & Shadlen, M. N. (2010). Temporal context calibrates interval timing. *Nature Neuroscience*, *13*(8), 1020–1026. https://doi.org/10.1038/nn.2590

Jiang, L. P., & Rao, R. P. N. (2024). Dynamic predictive coding: A model of hierarchical sequence learning and prediction in the neocortex. *PLOS Computational Biology*, *20*(2), e1011801. https://doi.org/10.1371/journal.pcbi.1011801

Jones, C. R. G., Malone, T. J. L., Dirnberger, G., Edwards, M., & Jahanshahi, M. (2008). Basal ganglia, dopamine and temporal processing: Performance on three timing tasks on and off medication in Parkinson's disease. *Brain and Cognition*, *68*(1), 30–41. https://doi.org/10.1016/j.bandc.2008.02.121

Kahalani-Hodedany, R., Lev, M., Sagi, D., & Polat, U. (2024). Generalization in perceptual learning across stimuli and tasks. *Scientific Reports*, *14*(1), 24546. https://doi.org/10.1038/s41598-024-75710-9

Karmarkar, U. R., & Buonomano, D. V. (2003). Temporal Specificity of Perceptual Learning in an Auditory Discrimination Task. *Learning & Memory*, *10*(2), 141–147. https://doi.org/10.1101/lm.55503

Keller, A. J., Dipoppa, M., Roth, M. M., Caudill, M. S., Ingrosso, A., Miller, K. D., & Scanziani, M. (2020). A Disinhibitory Circuit for Contextual Modulation in Primary Visual Cortex. *Neuron*, *108*(6), 1181-1193.e8. https://doi.org/10.1016/j.neuron.2020.11.013

Kelly, D. H. (1962). Visual Responses to Time-Dependent Stimuli. *Journal of the Optical Society of America*, *52*(1), 89–95. https://doi.org/10.1364/JOSA.52.000089

Kersten, D., Mamassian, P., & Yuille, A. (2004). Object perception as Bayesian inference. *Annual Review of Psychology*, *55*, 271–304. https://doi.org/10.1146/annurev.psych.55.090902.142005

Knill, D. C., & Richards, W. (Eds.). (1996). *Perception as Bayesian Inference* (1st ed.). Cambridge University Press. https://doi.org/10.1017/CBO9780511984037

Knudsen, E. I. (2004). Sensitive periods in the development of the brain and behavior. *Journal of Cognitive Neuroscience*, *16*(8), 1412–1425. https://doi.org/10.1162/0898929042304796

Komori, M. (2016). Effects of Working Memory Capacity on Metacognitive Monitoring: A Study of Group Differences Using a Listening Span Test. *Frontiers in Psychology*, *7*. https://doi.org/10.3389/fpsyg.2016.00285

Kondat, T., Tik, N., Sharon, H., Tavor, I., & Censor, N. (2024). Distinct Neural Plasticity Enhancing Visual Perception. *The Journal of Neuroscience, 44*(36), e0301242024. https://doi.org/10.1523/JNEUROSCI.0301-24.2024

Kondrakiewicz, K., & Nawrocka, J. (2025). Brains are expensive, but cognition is often cheap. *Neuroscience & Biobehavioral Reviews*, *179*, 106450. https://doi.org/10.1016/j.neubiorev.2025.106450

Kuhlman, S. J., Olivas, N. D., Tring, E., Ikrar, T., Xu, X., & Trachtenberg, J. T. (2013). A disinhibitory microcircuit initiates critical-period plasticity in the visual cortex. *Nature*, *501*(7468), 543–546. https://doi.org/10.1038/nature12485

Lachenmayr, B. J., Kojetinsky, S., Ostermaier, N., Angstwurm, K., Vivell, P. M., & Schaumberger, M. (1994). The different effects of aging on normal sensitivity in flicker and light-sense perimetry. *Investigative Ophthalmology & Visual Science*, *35*(6), 2741–2748.

Laughlin, S. B., de Ruyter van Steveninck, R. R., & Anderson, J. C. (1998). The metabolic cost of neural information. *Nature Neuroscience*, *1*(1), 36–41. https://doi.org/10.1038/236

Lee, H.-K., & Whitt, J. L. (2015). Cross-modal synaptic plasticity in adult primary sensory cortices. *Current Opinion in Neurobiology*, *35*, 119–126. https://doi.org/10.1016/j.conb.2015.08.002

Lee, J. L., Denison, R., & Ma, W. J. (2023). Challenging the fixed-criterion model of perceptual decision-making. *Neuroscience of Consciousness*, *2023*(1), niad010. https://doi.org/10.1093/nc/niad010

Lensjø, K. K., Lepperød, M. E., Dick, G., Hafting, T., & Fyhn, M. (2017). Removal of Perineuronal Nets Unlocks Juvenile Plasticity Through Network Mechanisms of Decreased Inhibition and Increased Gamma Activity. *The Journal of Neuroscience, 37*(5), 1269–1283. https://doi.org/10.1523/JNEUROSCI.2504-16.2016

Levelt, C. N., & Hübener, M. (2012). Critical-period plasticity in the visual cortex. *Annual Review of Neuroscience*, *35*, 309–330. https://doi.org/10.1146/annurev-neuro-061010-113813

Li, H.-H., Sprague, T. C., Yoo, A. H., Ma, W. J., & Curtis, C. E. (2021). Joint representation of working memory and uncertainty in human cortex. *Neuron*, *109*(22), 3699-3712.e6. https://doi.org/10.1016/j.neuron.2021.08.022

Li, R. W., Ngo, C., Nguyen, J., & Levi, D. M. (2011). Video-Game Play Induces Plasticity in the Visual System of Adults with Amblyopia. *PLOS Biology*, *9*(8), e1001135. https://doi.org/10.1371/journal.pbio.1001135

Li, W., Piëch, V., & Gilbert, C. D. (2004). Perceptual learning and top-down influences in primary visual cortex. *Nature Neuroscience*, *7*(6), 651–657. https://doi.org/10.1038/nn1255

Li, W., Piëch, V., & Gilbert, C. D. (2008). Learning to Link Visual Contours. *Neuron*, *57*(3), 442–451. https://doi.org/10.1016/j.neuron.2007.12.011

Lisman, J. E., & Jensen, O. (2013). The Theta-Gamma Neural Code. *Neuron*, *77*(6), 1002–1016. https://doi.org/10.1016/j.neuron.2013.03.007

Lombardi, A., Wang, Q., Stüttgen, M. C., Mittmann, T., Luhmann, H. J., & Kilb, W. (2023). Recovery kinetics of short-term depression of GABAergic and glutamatergic synapses at layer 2/3 pyramidal cells in the mouse barrel cortex. *Frontiers in Cellular Neuroscience*, *17*. https://doi.org/10.3389/fncel.2023.1254776

Lu, Z.-L., & Dosher, B. A. (2022). Current directions in visual perceptual learning. *Nature Reviews Psychology*, *1*(11), 654–668. https://doi.org/10.1038/s44159-022-00107-2

Łuczak, A., & Sobolewski, A. (2005). Longitudinal changes in critical flicker fusion frequency: An indicator of human workload. *Ergonomics*, *48*(15), 1770–1792. https://doi.org/10.1080/00140130500241753

Manea, A. M. G., Maisson, D. J.-N., Voloh, B., Zilverstand, A., Hayden, B., & Zimmermann, J. (2024). Neural timescales reflect behavioral demands in freely moving rhesus macaques. *Nature Communications*, *15*(1), 2151. https://doi.org/10.1038/s41467-024-46488-1

Mankowska, N. D., Grzywinska, M., Winklewski, P. J., & Marcinkowska, A. B. (2022). Neuropsychological and Neurophysiological Mechanisms behind Flickering Light Stimulus Processing. *Biology*, *11*(12). https://doi.org/10.3390/biology11121720

Mankowska, N. D., Marcinkowska, A. B., Waskow, M., Sharma, R. I., Kot, J., & Winklewski, P. J. (2021). Critical Flicker Fusion Frequency: A Narrative Review. *Medicina, 57*(10), 1096. https://doi.org/10.3390/medicina57101096

Mankowska, N. D., Sharma, R. I., Marcinkowska, A. B., Kot, J., & Winklewski, P. J. (2025). Assessing the Relationship Between the Flicker Test and Cognitive Performance. *Biology*, *14*(11), 1469. https://doi.org/10.3390/biology14111469

Mankowska, N. D., Sharma, R. I., Marcinkowska, A. B., Winklewski, P. J., & Kot, J. (2026). Complex Relationship Between Critical Flicker Fusion Frequency and Established Cognitive Tests Unveiled by Hyperbaric Exposure. *Biology*, *15*(3), 242. https://doi.org/10.3390/biology15030242

Manza, P., Tomasi, D., Demiral, Ş. B., Shokri-Kojori, E., Lildharrie, C., Lin, E., Wang, G.-J., & Volkow, N. D. (2025). Neural basis for individual differences in the attention-enhancing effects of methylphenidate. *Proceedings of the National Academy of Sciences, 122*(13), e2423785122. https://doi.org/10.1073/pnas.2423785122

Mastropasqua, T., & Turatto, M. (2013). Perceptual Grouping Enhances Visual Plasticity. *PLOS ONE*, *8*(1), e53683. https://doi.org/10.1371/journal.pone.0053683

Mather, M., & Sutherland, M. R. (2011). Arousal-biased competition in perception and memory. *Perspectives on Psychological Science, 6*(2), 114–133. https://doi.org/10.1177/1745691611400234

McGovern, D. P., Burns, S., Hirst, R. J., & Newell, F. N. (2022). Perceptual training narrows the temporal binding window of audiovisual integration in both younger and older adults. *Neuropsychologia*, *173*, 108309. https://doi.org/10.1016/j.neuropsychologia.2022.108309

McKendrick, A. M., & Battista, J. (2013). Perceptual learning of contour integration is not compromised in the elderly. *Journal of Vision*, *13*(1), 5. https://doi.org/10.1167/13.1.5

Meck, W. H. (1996). Neuropharmacology of timing and time perception. *Cognitive Brain Research*, *3*(3–4), 227–242. https://doi.org/10.1016/0926-6410(96)00009-2

Mendonça, A. G., Drugowitsch, J., Vicente, M. I., DeWitt, E. E. J., Pouget, A., & Mainen, Z. F. (2020). The impact of learning on perceptual decisions and its implication for speed-accuracy tradeoffs. *Nature Communications*, *11*(1), 2757. https://doi.org/10.1038/s41467-020-16196-7

Miederer, I., Buchholz, H.-G., Rademacher, L., Eckart, C., Kraft, D., Piel, M., Fiebach, C. J., & Schreckenberger, M. (2025). Dopaminergic Mechanisms of Cognitive Flexibility: An [18F] Fallypride PET Study. *Journal of Nuclear Medicine, 66*(3), 405–409. https://doi.org/10.2967/jnumed.124.268317

Miller, G. A. (1956). The magical number seven, plus or minus two: Some limits on our capacity for processing information. *Psychological Review*, *63(2), 81–97.* https://doi.org/10.1037/h0043158

Murray, J. D., Bernacchia, A., Freedman, D. J., Romo, R., Wallis, J. D., Cai, X., Padoa-Schioppa, C., Pasternak, T., Seo, H., Lee, D., & Wang, X.-J. (2014). A hierarchy of intrinsic timescales across primate cortex. *Nature Neuroscience*, *17*(12), 1661–1663. https://doi.org/10.1038/nn.3862

Muth, T., Schipke, J. D., Brebeck, A. K., & Dreyer, S. (2023). Assessing Critical Flicker Fusion Frequency: Which Confounders? A Narrative Review. *Medicina, 59*(4). https://doi.org/10.3390/medicina59040800

Muth, T., Schipke, J. D., Brebeck, A. K., & Dreyer, S. (2026). Critical flicker fusion frequency: Confounders and caveats. European Journal of Applied Physiology, 126, 725–734. https://doi.org/10.1007/s00421-025-05935-7

Nakamura, M., Valerio, P., Bhumika, S., & Barkat, T. R. (2020). Sequential Organization of Critical Periods in the Mouse Auditory System. *Cell Reports*, *32*(8), 108070. https://doi.org/10.1016/j.celrep.2020.108070

Nardella, A., Rocchi, L., Conte, A., Bologna, M., Suppa, A., & Berardelli, A. (2014). Inferior parietal lobule encodes visual temporal resolution processes contributing to the critical flicker frequency threshold in humans. *PLOS ONE, 9*(6), e98948. https://doi.org/10.1371/journal.pone.0098948

Oei, A. C., & Patterson, M. D. (2013). Enhancing Cognition with Video Games: A Multiple Game Training Study. *PLOS ONE*, *8*(3), e58546. https://doi.org/10.1371/journal.pone.0058546

Pajevic, S., Basser, P. J., & Fields, R. D. (2014). Role of myelin plasticity in oscillations and synchrony of neuronal activity. *Neuroscience*, *276*, 135–147. https://doi.org/10.1016/j.neuroscience.2013.11.007

Pan, Y. (2020). Visual working memory modulates the temporal resolution of visual perception. *Quarterly Journal of Experimental Psychology*, *73*(11), 1719–1728. https://doi.org/10.1177/1747021820925799

Parkin, C., Kerr, J. S., & Hindmarch, I. (1997). The Effects of Practice on Choice Reaction Time and Critical Flicker Fusion Threshold. *Human Psychopharmacology: Clinical and Experimental*, *12*(1), 65–70. https://doi.org/10.1002/(SICI)1099-1077(199701/02)12:1<65::AID-HUP838>3.0.CO;2-W

Pascucci, D., Tanrikulu, Ö. D., Ozkirli, A., Houborg, C., Ceylan, G., Zerr, P., Rafiei, M., & Kristjánsson, Á. (2023). Serial dependence in visual perception: A review. *Journal of Vision*, *23*(1), 9. https://doi.org/10.1167/jov.23.1.9

Pirazzini, G., & Ursino, M. (2024). Modeling the contribution of theta-gamma coupling to sequential memory, imagination, and dreaming. *Frontiers in Neural Circuits*, *18*. https://doi.org/10.3389/fncir.2024.1326609

Polat, U. (2009). Making perceptual learning practical to improve visual functions. *Vision Research*, *49*(21), 2566–2573. https://doi.org/10.1016/j.visres.2009.06.005

Powers, A. R., Hillock, A. R., & Wallace, M. T. (2009). Perceptual training narrows the temporal window of multisensory binding. *The Journal of Neuroscience, 29*(39), 12265–12274. https://doi.org/10.1523/JNEUROSCI.3501-09.2009

Qin, Y., Ahmadlou, M., Suhai, S., Neering, P., de Kraker, L., Heimel, J. A., & Levelt, C. N. (2023). Thalamic regulation of ocular dominance plasticity in adult visual cortex. *eLife*, *12*, RP88124. https://doi.org/10.7554/eLife.88124

Raichle, M. E., & Gusnard, D. A. (2002). Appraising the brain's energy budget. *Proceedings of the National Academy of Sciences*, *99*(16), 10237–10239. https://doi.org/10.1073/pnas.172399499

Rammsayer, T. H. (1997). Are there dissociable roles of the mesostriatal and mesolimbocortical dopamine systems on temporal information processing in humans? *Neuropsychobiology*, *35*(1), 36–45. https://doi.org/10.1159/000119328

Rao, R. P., & Ballard, D. H. (1999). Predictive coding in the visual cortex: A functional interpretation of some extra-classical receptive-field effects. *Nature Neuroscience*, *2*(1), 79–87. https://doi.org/10.1038/4580

Ribic, A., Crair, M. C., & Biederer, T. (2019). Synapse-Selective Control of Cortical Maturation and Plasticity by Parvalbumin-Autonomous Action of SynCAM 1. *Cell Reports*, *26*(2), 381-393.e6. https://doi.org/10.1016/j.celrep.2018.12.069

Rubia, K., Halari, R., Christakou, A., & Taylor, E. (2009). Impulsiveness as a timing disturbance: Neurocognitive abnormalities in attention-deficit hyperactivity disorder during temporal processes and normalization with methylphenidate. *Philosophical Transactions of the Royal Society B: Biological Sciences*, *364*(1525), 1919–1931. https://doi.org/10.1098/rstb.2009.0014

Rupert, D. D., & Shea, S. D. (2022). Parvalbumin-Positive Interneurons Regulate Cortical Sensory Plasticity in Adulthood and Development Through Shared Mechanisms. *Frontiers in Neural Circuits*, *16*, 886629. https://doi.org/10.3389/fncir.2022.886629

Rushton, W. A. H. (1951). A theory of the effects of fibre size in medullated nerve. *The Journal of Physiology*, *115*(1), 101–122. https://doi.org/10.1113/jphysiol.1951.sp004655

Saari, J. C. (2000). Biochemistry of visual pigment regeneration: The Friedenwald lecture. *Investigative Ophthalmology & Visual Science*, *41*(2), 337–348.

Sagi, D. (2011). Perceptual learning in *Vision Research*. *Vision Research, 51*(13), 1552–1566. https://doi.org/10.1016/j.visres.2010.10.019

Sagi, D., & Tanne, D. (1994). Perceptual learning: Learning to see. *Current Opinion in Neurobiology*, *4*(2), 195–199. https://doi.org/10.1016/0959-4388(94)90072-8

Sala, G., Tatlidil, K. S., & Gobet, F. (2018). Video game training does not enhance cognitive ability: A comprehensive meta-analytic investigation. *Psychological Bulletin, 144*(2), 111–139. https://doi.org/10.1037/bul0000139

Salthouse, T. A. (2010). Influence of age on practice effects in longitudinal neurocognitive change. *Neuropsychology*, *24*(5), 563–572. https://doi.org/10.1037/a0019026

Samaha, J., & Postle, B. R. (2015). The Speed of Alpha-Band Oscillations Predicts the Temporal Resolution of Visual Perception. *Current Biology*, *25*(22), 2985–2990. https://doi.org/10.1016/j.cub.2015.10.007

Samaha, J., & Romei, V. (2024). Alpha-Band Frequency and Temporal Windows in Perception: A Review and Living Meta-analysis of 27 Experiments (and Counting). *Journal of Cognitive Neuroscience*, *36*(4), 640–654. https://doi.org/10.1162/jocn_a_02069

Sandbrink, K., & Summerfield, C. (2024). Modelling cognitive flexibility with deep neural networks. *Current Opinion in Behavioral Sciences*, *57*, 101361. https://doi.org/10.1016/j.cobeha.2024.101361

Scher, D., Purcell, D. G., & Caputo, S. J. (1985). Visual acuity at two phases of the menstrual cycle. *Bulletin of the Psychonomic Society*, *23*(2), 119–121. https://doi.org/10.3758/BF03329799

Schoups, A., Vogels, R., Qian, N., & Orban, G. (2001). Practising orientation identification improves orientation coding in V1 neurons. *Nature*, *412*(6846), 549–553. https://doi.org/10.1038/35087601

Seitz, A. R., Náñez, J. E., Holloway, S. R., & Watanabe, T. (2005). Visual experience can substantially alter critical flicker fusion thresholds. *Human Psychopharmacology: Clinical and Experimental*, *20*(1), 55–60. https://doi.org/10.1002/hup.661

Seitz, A. R., Náñez, J. E., Sr., Holloway, S. R., & Watanabe, T. (2006). Perceptual Learning of Motion Leads to Faster Flicker Perception. *PLOS ONE*, *1*(1), e28. https://doi.org/10.1371/journal.pone.0000028

Semedo, J. D., Jasper, A. I., Zandvakili, A., Krishna, A., Aschner, A., Machens, C. K., Kohn, A., & Yu, B. M. (2022). Feedforward and feedback interactions between visual cortical areas use different population activity patterns. *Nature Communications*, *13*(1), 1099. https://doi.org/10.1038/s41467-022-28552-w

Simoncelli, E. P., & Olshausen, B. A. (2001). Natural Image Statistics and Neural Representation. *Annual Review of Neuroscience*, *24,* 1193–1216. https://doi.org/10.1146/annurev.neuro.24.1.1193

Spitmaan, M., Seo, H., Lee, D., & Soltani, A. (2020). Multiple timescales of neural dynamics and integration of task-relevant signals across cortex. *Proceedings of the National Academy of Sciences, 117*(36), 22522–22531. https://doi.org/10.1073/pnas.2005993117

Sporns, O., & Edelman, G. M. (1993). Solving Bernstein’s problem: A proposal for the development of coordinated movement by selection. *Child Development*, *64*(4), 960–981.

Sterling, P., & Laughlin, S. (2015). *Principles of Neural Design*. MIT Press. https://mitpress.mit.edu/9780262534680/principles-of-neural-design/

Stevenson, R. A., Siemann, J. K., Schneider, B. C., Eberly, H. E., Woynaroski, T. G., Camarata, S. M., & Wallace, M. T. (2014). Multisensory Temporal Integration in Autism Spectrum Disorders. *Journal of Neuroscience*, *34*(3), 691–697. https://doi.org/10.1523/JNEUROSCI.3615-13.2014

Stickgold, R., & Walker, M. P. (2007). Sleep-dependent memory consolidation and reconsolidation. *Sleep Medicine, 8*(4), 331–343. https://doi.org/10.1016/j.sleep.2007.03.011

Teich, A. F., & Qian, N. (2010). V1 orientation plasticity is explained by broadly tuned feedforward inputs and intracortical sharpening. *Visual Neuroscience*, *27*(1–2), 57–73. https://doi.org/10.1017/S0952523810000039

Treisman, M. (1963). Temporal discrimination and the indifference interval. Implications for a model of the ‘internal clock’. *Psychological Monographs*, *77*(13), 1–31. https://doi.org/10.1037/h0093864

Treisman, M., Cook, N., Naish, P. L., & MacCrone, J. K. (1994). The internal clock: Electroencephalographic evidence for oscillatory processes underlying time perception. *The Quarterly Journal of Experimental Psychology Section A, 47*(2), 241–289. https://doi.org/10.1080/14640749408401112

Treisman, M., Faulkner, A., & Naish, P. L. (1992). On the relation between time perception and the timing of motor action: Evidence for a temporal oscillator controlling the timing of movement.

*The Quarterly Journal of Experimental Psychology Section A, 45*(2), 235–263. https://doi.org/10.1080/14640749208401326

Trepka, E., Spitmaan, M., Qi, X.-L., Constantinidis, C., & Soltani, A. (2024). Training-Dependent Gradients of Timescales of Neural Dynamics in the Primate Prefrontal Cortex and Their Contributions to Working Memory. *Journal of Neuroscience*, *44*(2). https://doi.org/10.1523/JNEUROSCI.2442-21.2023

Tschantz, A., Millidge, B., Seth, A. K., & Buckley, C. L. (2023). Hybrid predictive coding: Inferring, fast and slow. *PLOS Computational Biology*, *19*(8), e1011280. https://doi.org/10.1371/journal.pcbi.1011280

Turgeon, M., Lustig, C., & Meck, W. H. (2016). Cognitive Aging and Time Perception: Roles of Bayesian Optimization and Degeneracy. *Frontiers in Aging Neuroscience*, *8*, 102. https://doi.org/10.3389/fnagi.2016.00102

Tyler, C. W. (1985). Analysis of visual modulation sensitivity II Peripheral retina and the role of photoreceptor dimensions. *Journal of the Optical Society of America A*, *2*(3), 393. https://doi.org/10.1364/JOSAA.2.000393

Urai, A. E., de Gee, J. W., Tsetsos, K., & Donner, T. H. (2019). Choice history biases subsequent evidence accumulation. *eLife*, *8*, e46331. https://doi.org/10.7554/eLife.46331

van den Berg, R., Yoo, A. H., & Ma, W. J. (2017). Fechner's law in metacognition: A quantitative model of visual working memory confidence. *Psychological Review*, *124*(2), 197–214. https://doi.org/10.1037/rev0000060

Van Hedger, S. C., Heald, S. L. M., & Nusbaum, H. C. (2019). Absolute pitch can be learned by some adults. *PLOS ONE*, *14*(9), e0223047. https://doi.org/10.1371/journal.pone.0223047

Vani, P. R., Nagarathna, R., Nagendra, H. R., & Telles, S. (1997). Progressive increase in critical flicker fusion frequency following yoga training. *Indian Journal of Physiology and Pharmacology*, *41*(1), 71–74.

VanRullen, R. (2016). Perceptual Cycles. *Trends in Cognitive Sciences*, *20*(10), 723–735. https://doi.org/10.1016/j.tics.2016.07.006

Vroomen, J., Keetels, M., de Gelder, B., & Bertelson, P. (2004). Recalibration of temporal order perception by exposure to audio-visual asynchrony. *Cognitive Brain Research*, *22*(1), 32–35. https://doi.org/10.1016/j.cogbrainres.2004.07.003

Wallace, M. T., & Stevenson, R. A. (2014). The construct of the multisensory temporal binding window and its dysregulation in developmental disabilities. *Neuropsychologia*, *64*, 105–123. https://doi.org/10.1016/j.neuropsychologia.2014.08.005

Walsh, J. F., & Misiak, H. (1966). Diurnal variation of critical flicker frequency. *The Journal of General Psychology*, *75*(1st Half), 167–175. https://doi.org/10.1080/00221309.1966.9710363

Ward, M. M., Stone, S. C., & Sandman, C. A. (1978). Visual perception in women during the menstrual cycle. *Physiology & Behavior*, *20*(3), 239–243. https://doi.org/10.1016/0031-9384(78)90215-9

Whitt, J. L., Petrus, E., & Lee, H.-K. (2014). Experience-dependent homeostatic synaptic plasticity in neocortex. *Neuropharmacology*, *78*, 45–54. https://doi.org/10.1016/j.neuropharm.2013.02.016

Wiesel, T. N., & Hubel, D. H. (1963). Single-cell responses in striate cortex of kittens deprived of vision in one eye. *Journal of Neurophysiology*, *26*(6), 1003–1017. https://doi.org/10.1152/jn.1963.26.6.1003

Willems, C., & Martens, S. (2016). Time to see the bigger picture: Individual differences in the attentional blink. *Psychonomic Bulletin & Review*, *23*, 1289–1299. https://doi.org/10.3758/s13423-015-0977-2

Wolf, E., & Schraffa, A. M. (1964). Relationship Between Critical Flicker Frequency and Age in Flicker Perimetry. *Archives of Ophthalmology, 72*, 832–845. https://doi.org/10.1001/archopht.1964.00970020834020

Wutz, A., Muschter, E., van Koningsbruggen, M. G., Weisz, N., & Melcher, D. (2016). Temporal Integration Windows in Neural Processing and Perception Aligned to Saccadic Eye Movements. *Current Biology*, *26*(13), 1659–1668. https://doi.org/10.1016/j.cub.2016.04.070

Xin, W., Kaneko, M., Roth, R. H., Zhang, A., Nocera, S., Ding, J. B., Stryker, M. P., & Chan, J. R. (2024). Oligodendrocytes and myelin limit neuronal plasticity in visual cortex. *Nature*, *633*(8031), 856–863. https://doi.org/10.1038/s41586-024-07853-8

Xu, R., Church, R. M., Sasaki, Y., & Watanabe, T. (2021). Effects of stimulus and task structure on temporal perceptual learning. *Scientific Reports*, *11*(1), 668. https://doi.org/10.1038/s41598-020-80192-6

Yamahachi, H., Marik, S. A., McManus, J. N. J., Denk, W., & Gilbert, C. D. (2009). Rapid axonal sprouting and pruning accompany functional reorganization in primary visual cortex. *Neuron*, *64*(5), 719–729. https://doi.org/10.1016/j.neuron.2009.11.026

Yao, H., & Dan, Y. (2001). Stimulus timing-dependent plasticity in cortical processing of orientation. *Neuron*, *32*(2), 315–323. https://doi.org/10.1016/s0896-6273(01)00460-3

Zhang, W., & Luck, S. J. (2008). Discrete fixed-resolution representations in visual working memory. *Nature*, *453*(7192), 233–235. https://doi.org/10.1038/nature06860

Zhou, T., Náñez, J. E., Zimmerman, D., Holloway, S. R., & Seitz, A. (2016). Two Visual Training Paradigms Associated with Enhanced Critical Flicker Fusion Threshold. *Frontiers in Psychology*, *7, Article 1597.* https://doi.org/10.3389/fpsyg.2016.01597

Zhu, H., He, F., Zolotavin, P., Patel, S., Tolias, A., Luan, L., & Xie, C. (2025). Temporal coding carries more stable cortical visual representations than firing rate over time. *Nature Communications*, *16*. https://doi.org/10.1038/s41467-025-62069-2

Zhu, S., Oh, Y. J., Trepka, E. B., Chen, X., & Moore, T. (2024). Dependence of contextual modulation in macaque V1 on interlaminar signal flow. *eLife*, *13*, RP103255. https://doi.org/10.7554/eLife.103255

Zivony, A., & Lamy, D. (2022). What processes are disrupted during the attentional blink? An integrative review of event-related potential research. *Psychonomic Bulletin & Review*, *29*(2), 394–414. https://doi.org/10.3758/s13423-021-01973-2

**Critical Flicker Fusion Frequency As An Experience-Restricted Constraint On Visual Temporal Resolution: What Does And Does Not Change It**

# Electronic Supplementary Material

### Supplementary Table S1. Training and intervention studies on CFFF

Referenced in Sections 1.2 and 4.1. For each study we indicate whether the protocol falls within the provisionally admitted class (perceptual learning that drives early visual processing of transient signals; Section 1.2). Δ CFFF = direction/magnitude of change relative to baseline.

| Study | Population | Paradigm | Duration | Δ CFFF | Admitted class? | Note |
|---|---|---|---|---|---|---|
| Seitz et al. (2005, 2006) | Healthy adults (n = 5) | Motion-direction perceptual learning (Direction-Training Group) | 9 d (1 h/d) | +30% (range 21–54%); no asymptote reached by day 9 | Yes (strongest evidence) | Untrained control group (Flicker-Only/CFFT-Only, n = 5) showed only +5% (range 3–9%) over the same 9 days; a further PreTest-PostTest group (n = 3, tested only before/after) showed +22% (range 18–27%), ruling out repeated testing alone as the cause. |
| Zhou et al. (2016) | Healthy adults (n = 10; 4 male; age 20.70 ± 4.42) | ULTIMEYES (contrast-sensitivity training) | 8 d × 25 min | +1.36 Hz (19.85 → 21.20 Hz, ≈ +6.8%); differed from Sudoku control; group × session interaction $p = 0.08$ | Yes | Only training condition to differ significantly from the active control |
| | Healthy adults (n = 10; 2 male; age 24.91 ± 8.58) | Directional dot-motion training | 8 d × 45 min | +0.74 Hz (18.86 → 19.60 Hz, ≈ +3.9%); not significantly different from control | Yes | Change within the familiarisation range |
| | Healthy adults (n = 10; 5 male; age 18.56 ± 1.01) | Sudoku (active control) | 8 d × 30 min | No effect (n.s.) | No | Active control for the class-restriction prediction |
| Eisen-Enosh et al. (2023) | Adults with amblyopia (n = 6) | Flicker-based temporal training | ~150 min (5 sessions over 2.5 weeks) | +3.9 Hz (22.8 → 26.7 Hz, ≈ +17%); individual gains 2.4–4.7 Hz | Yes | Amblyopic eye reached fellow-eye levels; trained eye started below typical values |
| | Healthy adults (n = 6) | Flicker-based temporal training | ~150 min | No effect | Yes (no effect) | Consistent with gains depending on headroom; confounded with population |
| Lambourne & Tomporowski (2010) | Healthy adults (21 studies with measures during and 29 with measures after exercise) | Acute aerobic exercise (multiple protocols, meta-regression) | Single bouts; literature to December 2008 | Not reported for CFF; CFF contributed effects from five primary studies to pooled estimates across all task types (during exercise Δ = −0.14; after exercise Δ = 0.20) | No (state effect) | Task type was analysed as a moderator, but CFF was not analysed separately; listed for context only and not used as evidence for a CFF-specific effect. |
| Lambourne, Audiffren, & Tomporowski (2010) | Healthy young adults (n = 19; mean age 21.1 y) | Acute aerobic exercise (cycle ergometer, 90% ventilatory threshold) | 40 min exercise (CFF measured before, 5× during, 3× after); separate seated control session | During exercise, rise from ≈ 35.9 to ≈ 39.1 Hz (≈ +9%; read from Fig. 2), differing from rest at min 28 and 34 ($p = 0.009$, .008); no difference from rest 1, 15 or 30 min after exercise (1 min: $p = 0.39$) | No (state effect) | Condition × block interaction $F(8, 144) = 5.98$, $p < 0.001$; ICC in the rest session 0.99. Pre-exercise mean lower than the rest-session pretest (≈ 35.9 vs ≈ 37.4 Hz), so part of the rise may reflect session differences. Distinct from the meta-regression above. |
| Vani et al. (1997) | Healthy male adults (n = 18 yoga; n = 18 control) | Yoga training (asanas, pranayama, kriyas, meditation, devotional sessions) | 30 d | No change at day 10; +11.1% at day 20; +14.9% at day 30 (yoga group only) | No | Allocation not randomised; group × assessment interaction not significant ($F = 0.1$); mechanism, if any, unknown |
| Bovier et al. (2014) | Healthy young adults (n = 54 active; n = 10 placebo) | Lutein/zeaxanthin supplementation (randomised, double-blind, placebo-controlled) | 4 months | Active groups +0.74 Hz (26.60 → 27.34 Hz, $p = 0.01$); placebo +0.11 Hz (n.s.) | No (physiological) | Non-experiential intervention; small-field (1°, 660 nm) LED stimulus; small placebo group |
| Bernardi et al. (2007) | Healthy adults (n = 20) | Repeated testing (familiarisation), flicker perimetry | 1st → 2nd of 5 sessions (first 3 across 1–30 d; last 3 same day) | ≈ +3% | No | Task familiarisation; saturates after a few trials. |

Summary of this table: (1) cognitive training without temporal content (Sudoku control) produced no CFFF change; (2) among training paradigms, gains are reported only for perceptual learning that drives early visual processing of transient signals, most clearly for coherent motion temporally paired with task targets, and every such study is small; (3) flicker training raised the threshold in amblyopic eyes starting below typical values but not in normally sighted observers; (4) acute exercise raised CFF during exercise only, with no difference from rest within 1 min of stopping, and the yoga result does not support a training-specific effect; (5) macular carotenoid supplementation produced a small increase relative to placebo; (6) the ~3% repeated-testing effect (Bernardi et al., 2007) is usually attributed to familiarisation.

**Supplementary Table S2. Synthesis of CFFF stability and state-modulation data**

Referenced in Sections 1.2, 2.1.2, 2.4, and 4.1.

**(a) Stability / reliability**

| Parameter | Value | Population | Source |
|---|---|---|---|
| Test–retest reliability (ICC, portable device) | 0.91–0.97 | Patients with chronic liver disease (n = 153), not healthy controls | Vutien et al. (2023) |
| Reliability under optimal conditions | r = 0.95 | Healthy adults (n = 100 university students); 2 sessions one day apart, 40 trials/subject | Jensen (1983) |
| Test–retest reliability (r) | 0.67–0.95 (moderately high) | Healthy male volunteers (n = 48), retested within 8 h | Görtelmeyer & Wiemann (1982) |
| Cross-method correlation (staircase vs. constant stimuli) | r = 0.92 | 10 healthy adults | Eisen-Enosh et al. (2017) |
| Bland–Altman agreement | ±3.6 Hz (95% CI) | 10 healthy adults | Eisen-Enosh et al. (2017) |
| Between- vs. within-individual variance | ~80% between participants; ~10% session (rise of 0.9–1.3 Hz per session); 4–9% residual | Healthy adults (n = 88 total; n = 49 completed 3 retest sessions) | Haarlem et al. (2024) |
| Max between-individual difference; 95% PI | ~30 Hz; 21 Hz | Healthy adults (n = 88) | Haarlem et al. (2024) |
| Within-individual CV (38 days) | ~4% (44.0 ± 1.7 Hz) | Single male adult (N = 1 pilot study) | Muth et al. (2023) |
| Within-individual CV (85 days) | ~5% (43.4 ± 2.1 Hz) | Single male diver (N = 1) | Muth et al. (2026) |
| Familiarisation (learning) effect | ~3% (1st → 2nd session) | Healthy adults (n = 20) | Bernardi et al. (2007) |
| Stability after familiarisation | No further improvement | Healthy men (n = 11), tested during parabolic flights | Balestra et al. (2018) |
| Diurnal variation | No simple decline: group CFF rose over the day, peaking at 8 p.m.; only a minority of individuals (7/60) showed the classically cited declining pattern | Healthy college students (n = 60; 30 male, 30 female) | Walsh & Misiak (1966) |
| Seasonal variation | Linear trend + oscillation; individual differences | Healthy male adults (n = 23; age 28–60), followed for 12 months | Łuczak & Sobolewski (2005) |
| Sex differences (mean CFFF) | No reliable mean difference; greater test–retest variability in women (males n = 20 stable; females n = 29 increased ~1.6 Hz vs. ~0.4 Hz per session) | Healthy adults (n = 88; retest subsample n = 49) | Haarlem et al. (2024) |
| Stability across repeated cognitive testing | CFFF stable across 3 sessions; age correlated with fusion (not flicker) threshold specifically; CFFF–cognition correlations weak | Healthy adults (n = 36) | Mankowska et al. (2025) |
| Age-related decline | Linear over 9–86 years, no acceleration (−0.38 dB/decade ≈ −4.3% per decade in Hz)* | Healthy people (n = 130; age 9–86) | Lachenmayr et al. (1994) |

* Lachenmayr et al. (1994) defined CFF(dB) = 20·log10 CFF (Hz); a slope of −0.38 dB/decade therefore corresponds to a factor of about 0.957, i.e. about −4.3% per decade. Values are mean flicker sensitivity across the central visual field (automated flicker perimetry), not a single foveal threshold.

**(b) State-dependent modulation (current performance, not the underlying limit)**

Referenced in Section 2.1.2. The state effects listed here largely reverse when normal conditions are restored, although reversal was incomplete 30 min after surfacing in Balestra et al. (2012); they are treated as modulation of current performance rather than change in the underlying limit.

| Factor | Direction | Mechanism | Source | Note |
|---|---|---|---|---|
| Fatigue / time-on-task | ↓ | Reduced cortical arousal | Muth et al. (2023); Simonson & Brozek (1952) | Simonson & Brozek (1952) is a classic foundational review, not a single primary study with one N. |
| Acute aerobic exercise | ↑ (during exercise only) | Increased arousal | Lambourne et al. (2010); Lambourne & Tomporowski (2010) | In the primary study (n = 19) CFF did not differ from rest 1 min after exercise; see Table S1. |
| Sleep deprivation | ↓ | Reduced vigilance | Muth et al. (2023) | Drawn from the authors' narrative review of prior literature, not from new primary data. |
| Stimulants (e.g., amphetamine, caffeine) | ↑ | Increased cortical arousal | Hindmarch (1982); MacNab et al. (1985) | MacNab et al. (1985): n = 13 (7 male, 6 female); single 10 mg doses; amphetamine increased CFF, diazepam decreased it, largest effects at 2–3 h post-dose. Hindmarch (1982) synthesises results from several of the author's own studies. |
| Nitrogen narcosis (hyperbaric) | ↓ | CNS depression (GABAergic) | Balestra et al. (2012) | n = 20 male divers (age 30–40); CFFF rose to 104% of baseline at depth, fell to 93.5% shortly before ascent, and remained at only 96.3% of baseline 30 min after surfacing. |
| Hyperoxia | Variable (dose-dependent) | Oxidative/metabolic | Muth et al. (2026) | n = 28 divers; CFFF measured before/after 5 min of normobaric oxygen breathing; effect on CFFF was minor. |
| Microgravity | ↑ | Increased cerebral arousal | Balestra et al. (2018) | n = 11 healthy men; CFFF reached 106.9 ± 5.5% of baseline during 0 g. |
| Hypergravity | ↓ | Reduced arousal | Balestra et al. (2018) | Same study; CFFF fell to 94.5 ± 3.8% of baseline during 2 g. |
| Breathing-gas mixture (air, heliox, trimix) at 4 ATA | ↑ (modest, 3–5%) | Hyperbaric exposure / altered arousal | Mankowska et al. (2026) | n = 40 (20 female; age 19–46); CFFF and reaction time both rose modestly (3–5%, $p < 0.05$) at 4 ATA across all three breathing gases, with no differential effect of gas type. |
| Menstrual cycle (hormonal state) | Inconsistent direction across studies | Hormonal fluctuation (oestrogen/progesterone) | Ward et al. (1978); Scher et al. (1985) | Direction of effect differs by visual task type (detection vs. acuity); already discussed in the main text as an example of genuine inconsistency rather than a single reliable state effect. |

## References

Balestra, C., Lafere, P., & Germonpre, P. (2012). Persistence of critical flicker fusion frequency impairment after a 33 mfw SCUBA dive: Evidence of prolonged nitrogen narcosis? *European Journal of Applied Physiology*, *112*(12), 4063–4068. https://doi.org/10.1007/s00421-012-2391-z

Balestra, C., Machado, M.-L. L., Theunissen, S., Balestra, A., Cialoni, D., Clot, C., Besnard, S., Kammacher, L., Delzenne, J., Germonpré, P., & Lafère, P. (2018). Critical flicker fusion frequency: A marker of cerebral arousal during modified gravitational conditions related to parabolic flights. *Frontiers in Physiology*, *9*, 1403. https://doi.org/10.3389/fphys.2018.01403

Bernardi, L., Costa, V. P., & Oto Shiroma, L. (2007). Flicker perimetry in healthy subjects: Influence of age and gender, learning effect and short-term fluctuation. *Arquivos Brasileiros de Oftalmologia*, *70*(1), 91–99. https://doi.org/10.1590/S0004-27492007000100017

Bovier, E. R., Renzi, L. M., & Hammond, B. R. (2014). A double-blind, placebo-controlled study on the effects of lutein and zeaxanthin on neural processing speed and efficiency. *PLOS ONE, 9*(9), e108178. https://doi.org/10.1371/journal.pone.0108178

Eisen-Enosh, A., Farah, N., Burgansky-Eliash, Z., Polat, U., & Mandel, Y. (2017). Evaluation of Critical Flicker-Fusion Frequency Measurement Methods for the Investigation of Visual Temporal Resolution. *Scientific Reports*, *7*(1), 15621. https://doi.org/10.1038/s41598-017-15034-z

Eisen-Enosh, A., Farah, N., Polat, U., & Mandel, Y. (2023). Perceptual learning based on a temporal stimulus enhances visual function in adult amblyopic subjects. *Scientific Reports*, *13*(1), 7643. https://doi.org/10.1038/s41598-023-34421-3

Görtelmeyer, R., & Wiemann, H. (1982). Retest Reliability and Construct Validity of Critical Flicker Fusion Frequency. *Pharmacopsychiatry*, *15*, 24–28. https://doi.org/10.1055/s-2007-1019545

Haarlem, C., O'Connell, R., Mitchell, K., & Jackson, A. (2024). The speed of sight: Individual variation in critical flicker fusion thresholds. *PLOS ONE*, *19(4), e0298007.* https://doi.org/10.1371/journal.pone.0298007

Hindmarch, I. (1982). Critical Flicker Fusion Frequency (CFF): The Effects of Psychotropic Compounds. *Pharmacopsychiatry*, *15*, 44–48. https://doi.org/10.1055/s-2007-1019549

Jensen, A. R. (1983). Critical flicker frequency and intelligence. *Intelligence*, *7*(2), 217–225. https://doi.org/10.1016/0160-2896(83)90030-2

Lachenmayr, B. J., Kojetinsky, S., Ostermaier, N., Angstwurm, K., Vivell, P. M., & Schaumberger, M. (1994). The different effects of aging on normal sensitivity in flicker and light-sense perimetry. *Investigative Ophthalmology & Visual Science*, *35*(6), 2741–2748.

Lambourne, K., Audiffren, M., & Tomporowski, P. D. (2010). Effects of acute exercise on sensory and executive processing tasks. *Medicine and Science in Sports and Exercise*, *42*(7), 1396–1402. https://doi.org/10.1249/MSS.0b013e3181cbee11

Lambourne, K., & Tomporowski, P. (2010). The effect of exercise-induced arousal on cognitive task performance: A meta-regression analysis. *Brain Research, 1341*, 12–24. https://doi.org/10.1016/j.brainres.2010.03.091

Łuczak, A., & Sobolewski, A. (2005). Longitudinal changes in critical flicker fusion frequency: An indicator of human workload. *Ergonomics*, *48*(15), 1770–1792. https://doi.org/10.1080/00140130500241753

MacNab, M. W., Foltz, E. L., & Sweitzer, J. (1985). Evaluation of signal detection theory on the effects of psychotropic drugs on critical flicker-fusion frequency in normal subjects. *Psychopharmacology*, *85*(4), 431–435. https://doi.org/10.1007/BF00429659

Mankowska, N. D., Sharma, R. I., Marcinkowska, A. B., Kot, J., & Winklewski, P. J. (2025). Assessing the Relationship Between the Flicker Test and Cognitive Performance. *Biology*, *14*(11), 1469. https://doi.org/10.3390/biology14111469

Mankowska, N. D., Sharma, R. I., Marcinkowska, A. B., Winklewski, P. J., & Kot, J. (2026). Complex Relationship Between Critical Flicker Fusion Frequency and Established Cognitive Tests Unveiled by Hyperbaric Exposure. *Biology*, *15*(3), 242. https://doi.org/10.3390/biology15030242

Muth, T., Schipke, J. D., Brebeck, A. K., & Dreyer, S. (2023). Assessing Critical Flicker Fusion Frequency: Which Confounders? A Narrative Review. *Medicina, 59*(4). https://doi.org/10.3390/medicina59040800

Muth, T., Schipke, J. D., Brebeck, A. K., & Dreyer, S. (2026). Critical flicker fusion frequency: Confounders and caveats. European Journal of Applied Physiology, 126, 725–734. https://doi.org/10.1007/s00421-025-05935-7

Scher, D., Purcell, D. G., & Caputo, S. J. (1985). Visual acuity at two phases of the menstrual cycle. *Bulletin of the Psychonomic Society*, *23*(2), 119–121. https://doi.org/10.3758/BF03329799

Seitz, A. R., Náñez, J. E., Holloway, S. R., & Watanabe, T. (2005). Visual experience can substantially alter critical flicker fusion thresholds. *Human Psychopharmacology: Clinical and Experimental*, *20*(1), 55–60. https://doi.org/10.1002/hup.661

Simonson, E., & Brozek, J. (1952). Flicker Fusion Frequency: Background and Applications. *Physiological Reviews*, *32*(3), 349–378. https://doi.org/10.1152/physrev.1952.32.3.349

Vani, P. R., Nagarathna, R., Nagendra, H. R., & Telles, S. (1997). Progressive increase in critical flicker fusion frequency following yoga training. *Indian Journal of Physiology and Pharmacology*, *41*(1), 71–74.

Vutien, P., Li, R., Karkar, R., Munson, S. A., Fogarty, J., Walter, K., Yacoub, M., & Ioannou, G. N. (2023). Evaluating a Novel, Portable, Self-Administrable Device (‘Beacon’) That Measures Critical Flicker Frequency as a Test for Hepatic Encephalopathy. *The American Journal of Gastroenterology*, *118*(6), 1096–1100. https://doi.org/10.14309/ajg.0000000000002211

Walsh, J. F., & Misiak, H. (1966). Diurnal variation of critical flicker frequency. *The Journal of General Psychology*, *75*(1st Half), 167–175. https://doi.org/10.1080/00221309.1966.9710363

Ward, M. M., Stone, S. C., & Sandman, C. A. (1978). Visual perception in women during the menstrual cycle. *Physiology & Behavior*, *20*(3), 239–243. https://doi.org/10.1016/0031-9384(78)90215-9

Zhou, T., Náñez, J. E., Zimmerman, D., Holloway, S. R., & Seitz, A. (2016). Two Visual Training Paradigms Associated with Enhanced Critical Flicker Fusion Threshold. *Frontiers in Psychology*, *7, Article 1597.* https://doi.org/10.3389/fpsyg.2016.01597